\documentclass[11pt,draftcls,onecolumn]{IEEEtran}
\usepackage{etex} 
\usepackage{amsmath,graphicx,amssymb,epsfig,setspace,multicol,cite}
\usepackage{multirow}
\usepackage{float,dsfont}
\usepackage[latin1]{inputenc}
\usepackage{graphicx}
\usepackage{mathtools}
\usepackage{ulem} 
\usepackage{bbm} 
\usepackage{schemabloc}  
\usepackage{epsfig}       
\usepackage{pst-all}      
\usepackage{cancel}

\usepackage{tikz}
\usepackage{epstopdf}
\usetikzlibrary{positioning}

\DeclareTextSymbol{\degre}{T1}{6}
\DeclareTextSymbol{\degre}{OT1}{23}

\newcommand\tra{{\rm Tr}}

\newcommand\cov{{\rm Cov}}

\newcommand\e{{\rm E}}
\newcommand\pardef{ \stackrel{{\rm def}}{=} }

\newtheorem{result}{Result}

\def\QED{\mbox{\rule[0pt]{1.4ex}{1.4ex}}}

\usepackage{soul} 
\usepackage{color}

\begin{document}

\title{Asymptotic Frequency-Domain FIM for multidimensional
stationary and cyclostationary compound-Gaussian processes: Application to DOA estimation
}
\author{Jean-Pierre Delmas and Habti Abeida
\thanks{
Jean-Pierre Delmas is with Samovar laboratory, Telecom SudParis, Institut Polytechnique de Paris, 91120 Palaiseau, France,
e-mail: jean-pierre.delmas@it-sudparis.eu.
Habti Abeida is with Taif University, College of Engineering, Dept. of  Electrical Engineering, Al-Haweiah, 21974, Saudi Arabia,
email: h.abeida@tu.edu.sa,
}}
\maketitle

%
\begin{abstract}
The classical Whittle formula provides a closed-form expression for the asymptotic Fisher information matrix (FIM) of stationary, zero-mean Gaussian processes (GPs) with finite-dimensional parameterizations, expressed in terms of their power spectra. While this result was recently extended to nonzero-mean, multidimensional, real-valued stationary compound-Gaussian processes (CGPs), its applicability remains limited. In this paper, we significantly broaden this theoretical framework with two primary contributions. First, we extend the Whittle formula to purely cyclostationary real-valued CGPs with parameterized cyclic spectra,  and further generalize it to circular and noncircular complex-valued multidimensional stationary CGPs, encompassing both  purely and almost cyclostationary CGPs. Second, we apply these novel Whittle-type formulas  to evaluate the impact of source spectral characteristics on the Cram\'er-Rao bound (CRB) for direction-of-arrival (DOA) estimation, particularly in scenarios involving frequency offsets.
\end{abstract}
%

\maketitle
\centerline{
Submitted to Signal Processing
}
\date{\today}
%
\section{Introduction}
\label{sec:Introduction}
%
Spherically invariant random processes (SIRPs), also known as compound-Gaussian processes (CGPs), provide a flexible framework for modeling a broad class of non-Gaussian phenomena encountered in signal processing. This class includes, among others, $\epsilon$-contaminated Gaussian, Student-$t$, and $K$-distributed processes, as well as certain symmetric $\alpha$-stable processes, while also encompassing Gaussian processes as a particular case. The simplest models in this class are second-order (wide-sense) stationary parametric processes, for which the parameters of interest are estimated from a finite set of consecutive observations.

Evaluating the accuracy of parametric estimates in such  models is fundamental to the analysis and design of signal processing systems. This accuracy is commonly characterized by the Cram\'er-Rao bound (CRB), which is determined by the inverse of the Fisher information matrix  (FIM) and provides a lower bound on the covariance matrix of unbiased estimators. While the finite-sample FIM is given by the Slepian-Bangs formula \cite{Slepian1954,Bangs1971}, evaluating it is computationally prohibitive for large data records. In contrast, Whittle's formula \cite{Whittle1953} provides a closed-form expression for the asymptotic FIM rate of zero-mean, real-valued stationary Gaussian processes in terms of their power spectra.  This drastically reduces computational complexity while offering a direct interpretation in terms of the spectral properties of the process. Recently, by exploiting the asymptotic properties of block Toeplitz matrices, this result was extended in \cite{Delmas2026} to real-valued, arbitrary-mean stationary SIRPs within a unified framework covering fully known, finite-dimensional parameterized, and completely unspecified density generators.

However, several important extensions are required to make such asymptotic FIM results applicable to a broader class of signal processing problems. 
First, many signals are frequency down-converted to baseband after filtering and are consequently represented by their complex-valued envelopes, whose distributions may be either circular or noncircular.
Second, many man-made signals are not stationary but exhibit periodic statistical properties and are therefore naturally described by cyclostationary models. Such models arise in communication systems \cite{Serpedina2005,Napolitano2016}, rotating machinery \cite{Sun2020,Antoni2004}, radar and sonar \cite{Gini2000,Zhang2017}, and power-line communications \cite{Chien2020,Moaveninejad2020}. Third, practical communication signals may exhibit almost cyclostationary behavior, for which the relevant second-order statistics vary periodically with time but do not necessarily admit the strict periodicity of conventional cyclostationary models. A general asymptotic FIM framework encompassing these cases is therefore of considerable interest.

In this paper, we develop a unified extension of Whittle's formula to multidimensional stationary, cyclostationary, and almost cyclostationary CGPs. The main contributions are as follows.
\begin{itemize}
\item We derive the asymptotic frequency-domain FIM rate for {\it real-valued cyclostationary CGPs} with parameterized cyclic spectra as a function of the poly-spectral matrix, thereby extending the stationary result of \cite{Delmas2026} to processes with periodic second-order statistics.

\item We generalize the framework to {\it multidimensional complex-valued stationary CGPs}, covering both circular and noncircular processes. The latter case requires accounting for the complementary second-order statistics in addition to the conventional covariance structure.

\item We further extend the results to {\it purely and specifically almost cyclostationary complex-valued CGPs}. In particular, we establish results that enable the treatment of linearly modulated digital communication signals exhibiting cyclostationarity and their extension to specific multidimensional almost cyclostationary models.

\item We apply the resulting asymptotic FIM expressions to {\it DOA estimation} and characterize the impact of the source spectral characteristics on the corresponding CRB. Both stationary and cyclostationary source signals are considered, including scenarios with and without frequency offsets.
\end{itemize}

The proposed framework goes beyond the commonly used performance analysis based on independent and identically distributed Gaussian observations. In the considered setting, the observations are samples of continuous-time stochastic processes whose temporal and spectral characteristics can be arbitrary and whose distributions need not be Gaussian. Consequently, the resulting CRB explicitly incorporates information carried by the temporal spectrum, cyclic spectrum, and, for noncircular complex processes, complementary spectral characteristics. This provides a general analytical framework for studying how signal statistics influence parameter-estimation performance.

The remainder of this paper is organized as follows. Section \ref{sec:Background statistical} introduces the required preliminaries on real-valued multidimensional stationary CGPs and reviews the asymptotic frequency-domain FIM rate formula for these processes. Section \ref{sec:Asymptotic FIM rate for cyclostationary CGP} extends this formula to real-valued cyclostationary CGPs. Section \ref{sec:Asymptotic FIM rate for stationary, cyclostationary and almost stationary complex-valued CGP} further generalizes the results  to circular and noncircular complex-valued multidimensional stationary CGPs, as well as to purely and specifically almost cyclostationary CGPs. Preliminary results for linearly modulated digital communication signals exhibiting cyclostationarity are also established. Section \ref{sec:Application to DOA estimation} applies the proposed asymptotic FIM rate formulas to investigate the influence of source spectral characteristics on the CRB for DOA estimation, with particular emphasis on measurements affected by frequency offsets. Finally,  Section  \ref{sec:Conclusion}  concludes the paper.

The following notations are used throughout this paper.
Matrices
and vectors are represented by bold upper-case and bold lower-case
characters, respectively. Vectors are by default in column
orientation, while the superscripts $T$, $H$ and $*$ stand for transpose,
conjugate transpose and conjugate, respectively.
$\e(.)$, $|.|$, and $\tra(.)$ are the expectation, determinant, and trace operators, respectively.
$\|.\|_F$ denotes the Frobenius norm.
${\bf I}$ is the identity matrix with the appropriate dimension.
$\otimes$ is the Kronecker product of two matrices.
${\rm Blkdiag}({\bf B}_1,...,{\bf B}_n)$ is the block diagonal matrix with
${\bf B}_1,...,{\bf B}_n$
along its diagonal.
$\mathds{1}_{C}(f) \!=\! 1$ if $f \in C$ and $\mathds{1}_{C}(f) \!=\! 0$ if $f \not\in C$
and $x \sim P_x$ means that the random variables $x$ has the distribution $P_x$.
The acronyms r.v., p.d.f., c.d.f. and i.f.f. are used to represent random variable, probability density function, cumulative distribution function, and ''if and only if'', respectively.
%
\section{Statistical Background}
\label{sec:Background statistical}
%
\subsection{Preliminaries on multidimensional CGP}
\label{sec:Preliminaries on multidimensional CGP}
%
Let $({\bf x}_{k})_{k \in \mathbb{Z}}$ be a stochastic process with ${\bf x}_{k}\in \mathbb{R}^m$. This process is defined as an $m$-dimensional SIRP, or equivalently a  CGP, if any finite collection of samples ${\bf y} \pardef ({\bf x}_{k_1}^T,..,{\bf x}_{k_n}^T)^T$ with $k_1<...<k_n$, follows a  compound-Gaussian distribution. The  p.d.f. is given by:
\begin{equation}
\label{eq:pdf RES}
p({\bf y})
=
|\boldsymbol{\Sigma}|^{-1/2}g_{nm}[({\bf y}-\boldsymbol{\mu})^T \boldsymbol{\Sigma}^{-1}({\bf y}-\boldsymbol{\mu})].
\end{equation}
The first and second-order moments of ${\bf y}$ are assumed to exist, satisfying $\e({\bf y})=\boldsymbol{\mu}$ and
\begin{equation}
\label{eq:cov}
\cov({\bf y})=\boldsymbol{\Sigma}
\end{equation}
respectively. This normalization in \eqref{eq:cov} is imposed to remove any scale ambiguity between $\boldsymbol{\Sigma}$ and $g_{nm}(.)$. The density generator $g_{nm}(.)$ is given by
\begin{equation}
\label{eq:density generator CG}
g_{nm}(t)
=
(2\pi)^{-nm/2}
\int_0^{\infty}
\tau^{-nm/2}\exp(-t/2\tau)
dF_{\tau}(\tau),
\end{equation}
where $F_{\tau}$ denotes the c.d.f. of the r.v. $\tau$, commonly referred to as the texture variable. According to \cite[Th. 2.2]{Yao1973}, the p.d.f. in \eqref{eq:pdf RES}, with the density generator given by \eqref{eq:density generator CG}, is equivalently induced by the stochastic representation
\begin{equation}
\label{eq:Stochastic representation CG}
{\bf y}
={\boldsymbol \mu}+ \sqrt{\tau}{\boldsymbol{\Sigma}}^{1/2}{\bf n}_0,
\end{equation}
where $\tau$ and ${\bf n}_0$ are independent, with ${\bf n}_0\sim{\cal N}_{nm}({\bf 0},{\bf I})$. Under the normalization condition in \eqref{eq:cov}, the texture variable satisfies $\e(\tau)=1$.
The density generator $g_{nm}(\cdot)$, or equivalently the c.d.f.  $F_{\tau}$ of $\tau$, may be fully known, parameterized by a finite-dimensional nuisance parameter \cite{Abeida2023}, or {completely unspecified} \cite{Fortunati2026}.

In the following, the above definition is applied to stationary and cyclostationary processes, with the covariance matrix $\boldsymbol{\Sigma}$ and the corresponding density generator determined by the temporal structure of the sampled observations.
%
\subsection{Preliminaries on multidimensional stationary CGP}
\label{sec:Preliminaries on multidimensional stationary CGP}
%
A CGP is said to be strictly stationary if the p.d.f. in \eqref{eq:pdf RES} of any finite collection of samples ${\bf y}$ is invariant under arbitrary translations of the sampling indices $(k_1,\ldots,k_n)$. Equivalently, the \r.v.s $({\bf x}_{1+k},\ldots,{\bf x}_{n+k})$ and $({\bf x}_{1},\ldots,{\bf x}_{n})$ share the same joint distribution for all $n\in\mathbb{N}^*$ and $k\in\mathbb{Z}$.
As for Gaussian processes, the compound-Gaussian structure implies that stationarity is equivalent to wide-sense stationarity.
More precisely, a CGP is stationary if and only if $\e({\bf x}_{\ell})$ and $\e({\bf x}_{\ell}{\bf x}_{\ell+k}^{T})$ are independent of $\ell\in\mathbb{Z}$ for every $k\in\mathbb{Z}$.
Consequently, for a stationary CGP, the random vector formed by stacking $n$ consecutive observations of ${\bf x}k$ is given by
\begin{equation}
\label{eq:yn}
{\bf y}_n\pardef({\bf x}_{1}^T,...,{\bf x}_{n}^T)^T \in \mathbb{R}^{mn}\ \
\forall  n \in \mathbb{N}^*,
\end{equation}
has mean
${\boldsymbol \mu}_{{\bf y}_n} \pardef\e({\bf y}_{n})
=(\mathds{1}_{n}\otimes {\bf I}_{m}){\boldsymbol \mu}_{x}$ with 
${\boldsymbol \mu}_{x}\pardef \e({\bf x}_{\ell})$ and covariance matrix $\boldsymbol{\Sigma}_{{\bf y}_n} \pardef \cov({\bf y}_n)$,
which is symmetric positive definite and has a block Toeplitz structure with block size $m$.
Its $(i,j)$-th block is given by ${\bf R}_{x}(j-i)$ with 
${\bf R}_{x}(k) \pardef \e[({\bf x}_{\ell}-{\boldsymbol \mu}_x)({\bf x}_{\ell+k}-{\boldsymbol \mu}_x)^T] \in \mathbb{R}^{m \times m}$.
We assume that
the sequence ${\bf R}_{x}(k)$ is absolutely summable, i.e., $\sum_{k=-\infty}^{\infty}\|{\bf R}_{x}(k)\|_2<\infty$, which 
guarantees that the spectrum\footnote{All Fourier transforms of sequences, whether scalar or matrix-valued, are denoted using the same notation, with the time index $k$ replaced by the frequency variable $f$.} ${\bf R}_{x}(f) \pardef \sum_k {\bf R}_{x}(k)e^{-i2\pi kf}$ is well-defined.
%
\subsection{Asymptotic Frequency-Domain FIM Rate for Stationary CGP}
\label{sec:Asymptotic FIM rate for stationary CGP}
%
Suppose that the mean ${\boldsymbol \mu}_x$ and the sequence ${\bf R}_{x}(k)$ 
depend on a parameter $\boldsymbol{\theta}=(\theta_1,..,\theta_r)^T \in \boldsymbol{\Theta}  \subset \mathbb{R}^r$, and consequently, so does the spectrum
${\bf R}_{x}(f)$.  For notational simplicity, this dependence on $\boldsymbol{\theta}$ is omitted whenever no ambiguity arises.
Assume that the mapping $\boldsymbol{\theta} \mapsto ({\boldsymbol \mu}_x,[{\bf R}_{x}(k)]_{k\in \mathbb{Z}},{\bf R}_{x}(f))$ is differentiable in a neighborhood of the true parameter value and that the regularity conditions stated in \cite{Delmas2026} hold, among which is $\e(\tau^{-1})<\infty$. Then, combining the Slepian-Bangs expression for the finite-sample FIM with the asymptotic theory of block Toeplitz matrices yields
the following asymptotic frequency domain FIM rate for a multidimensional  real-valued stationary CGP:
\begin{eqnarray}
\nonumber
\lim_{n \rightarrow \infty}
\frac{1}{n}\left({\rm FIM}_{{\bf y}_n}(\boldsymbol{\theta})\right)_{k,\ell}
&=&
\e\left(\tau^{-1}\right) {\boldsymbol \mu}^{'T}_{x,k}
{\bf R}_{x}^{-1}(0)
{\boldsymbol \mu}_{x,\ell}'
\\
\nonumber
&+&
\frac{1}{2}\int_{0}^{1}\tra({\bf R}_{x}^{-1}(f) {\bf R}_{x,k}^{'}(f) {\bf R}_{x}^{-1}(f){\bf R}_{x,\ell}^{'}(f))df
\\
\label{eq:Whittle's formula}
&+&
c
\left(\int_{0}^{1}\tra({\bf R}_{x}^{-1}(f) {\bf R}_{x,k}^{'}(f))df\right)
\left(\int_{0}^{1}\tra({\bf R}_{x}^{-1}(f) {\bf R}_{x,\ell}^{'}(f))df\right),
\end{eqnarray}
for $k,\ell=1,\ldots,r$, where ${\boldsymbol \mu}'_{x,k}\pardef \frac{\partial {\boldsymbol \mu}_{x}}{\partial \theta_k}$,
${\bf R}_{x}(0)={\bf R}_{x}(f)_{|f=0}$ denotes the spectrum matrix evaluated at zero frequency, 
${\bf R}_{x,k}^{'}(f)\pardef \frac{\partial {\bf R}_{x}(f)}{\partial \theta_k}$,
and 
where $c$ is given by \eqref{eq:lim a2} or \eqref{eq:lim a2 SePa}
depending on  whether the $g_{nm}(\cdot)$, or equivalently the c.d.f.  $F_{\tau}$ of $\tau$ is fully known or completely unspecified, respectively.
For the classical parametric case in which $F_\tau$ is known, the coefficient $c$ is
\begin{equation}
\label{eq:lim a2}
c_{\rm Clas}
=
-\frac{1-\beta}{2m}
\leq 0,
\end{equation}
where $\beta$ denotes the total probability mass of the discrete component of the distribution of $\tau$, i.e., $\beta \pardef \sum_{t:\,P(\tau=t)>0}P(\tau=t).$
For the semiparametric case in which the density generator is completely unspecified,
\begin{equation}
\label{eq:lim a2 SePa}
c_{\rm SePa}
=-\frac{1}{2m}.
\end{equation}
Eq. \eqref{eq:Whittle's formula} extends Whittle's frequency-domain FIM rate to multidimensional stationary CGPs with arbitrary mean. In the Gaussian case, 
$\tau=1$ and thus $c_{\rm Clas}=0$, recovering the multidimensional Gaussian Whittle formula with arbitrary mean.
%
\section{Asymptotic FIM rate for cyclostationary CGP}
\label{sec:Asymptotic FIM rate for cyclostationary CGP}
%
\subsection{Cyclostationary CGP with cycle period $p$}
\label{sec:Cyclostationary CGP with cycle period p}
%
Analogous to strict stationarity, a CGP is strictly cyclostationary with cycle period $p$ if its finite-dimensional distributions are invariant under time shifts of $p$.
This means that  r.v.s $({\bf x}_{1+pk},\dots,{\bf x}_{n+pk})$ and $({\bf x}_{1},\dots,{\bf x}_{n})$ share the same joint distribution for all $n\in \mathbb{N}^*$, $k \in \mathbb{Z}$. Within the standard SIRP framework, a CGP is strictly cyclostationary if and only if it is wide-sense
cyclostationary.\footnote{For a rigorous treatment of cyclostationary processes and a summary of their key properties, see 
\cite{Gardner1994,Napolitano2012,Napolitano2016}}. Precisely, a CGP is cyclostationary with cycle period $p$ i.f.f.
\begin{equation}
\label{eq:def wide sense cyclostationary}
\e({\bf x}_{k+ ip}) 
\ \mbox{and}\ 
\e({\bf x}_{k+ ip}{\bf x}_{\ell+ ip}^T) 
\ \mbox{are independent of}\ i\in \mathbb{Z},\ \forall k,l \in \mathbb{Z}.
\end{equation}
For a cyclostationary CGP with cycle period $p$,
the blocked sequence $({\bf x}_k^{'})_{k \in \mathbb{Z}}\in  \mathbb{R}^{mp}$, $k\in \mathbb{Z}$ with
${\bf x}^{'}_{k} \pardef [{\bf x}_{1+(k-1)p}^T,...,{\bf x}_{p+(k-1)p}^T]^T$ is stationary as a direct consequence of  \eqref{eq:def wide sense cyclostationary}
(see also \cite{Gladyshev1961},
\cite{Nematollahi2005}).

When the number of observations $n$ is a multiple of $p$  (i.e., $n = pq$ for some integer $q$) \cite{Raminez2015},  the  
vector ${\bf y}_n$ defined in \eqref{eq:yn} can be partitioned into $q$ blocks of size $mp$:
%
\begin{equation}
\label{eq:yn cyclostat}
{\bf y}_n=(\underbrace{{\bf x}_{1}^T,...,{\bf x}_{p}^T}_{{\bf x}_{1}^{'T}},...,\underbrace{{\bf x}_{(k-1)p+1}^T,...,{\bf x}_{kp}^T}_{{\bf x}_{k}^{'T}},...,
\underbrace{{\bf x}_{n-p+1}^T,...,{\bf x}_{n}^T}_{{\bf x}_{q}^{'T}})^T
\in \mathbb{R}^{mn}\
\mbox{for}\ n=pq.
\end{equation}
Consequently, the mean of ${\bf y}_n$ is given by
${\boldsymbol \mu}_{{\bf y}_n} =(\mathds{1}_{q}\otimes {\bf I}_{m}){\boldsymbol \mu}_{x'}$ with
${\boldsymbol \mu}_{x'} \pardef({\boldsymbol \mu}_{x_1}^T,...,{\boldsymbol \mu}_{x_p}^T)^T$. The covariance
matrix $\boldsymbol{\Sigma}_{{\bf y}_n} \pardef \mathrm{Cov}({\bf y}_n)$  is also a
symmetric positive definite and possesses a block-Toeplitz structure with block size $m p$. its $(k,\ell)$-th block is given by
${\bf R}_{x'}(\ell-k)$ for which the sequence
${\bf R}_{x'}(k) \pardef \e[({\bf x}'_{\ell} - \boldsymbol{\mu}_{x'})({\bf x}'_{\ell+k} - \boldsymbol{\mu}_{x'})^T] \in \mathbb{R}^{pm \times pm}$
is assumed to be absolutely summable. The corresponding spectral matrix of the blocked
$({\bf x}_k^{'})_{k \in \mathbb{Z}}\in  \mathbb{R}^{mp}$ is ${\bf R}_{x'}(f) \pardef \sum_{k\in\mathbb{Z}}{\bf R}_{x'}(k)e^{-i2\pi kf}$. Notably, the second-order frequency properties of the cyclostationary CGP are more advantageously
characterized by its cyclic spectra\footnote{{ ${\bf R}_{x'}(f)$ as spectrum in the conventional sense of  ${\bf x}'_k$ is defined by $p^2m^2$ scalar complex-valued functions, while the cyclic matrices ${\bf R}_{x}^{(\frac{k}{p})}(f)$, $k=0,..,p-1$ are only defined by  $p m^2$ scalar complex-valued functions which have generally more insightful interpretations.}}.
The cyclic spectral matrices are defined as:
${\bf R}_{x}^{(\frac{k}{p})}(f)\pardef \sum_{\tau=-\infty}^{\infty}{\bf R}_{x}^{(\frac{k}{p})}(\tau)e^{-i2\pi \tau f}$, $k=0,1,..,p-1$, where the cyclic autocovariance sequence
${\bf R}_{x}^{(\frac{k}{p})}(\tau)
\pardef
\frac{1}{p}\sum_{\ell=0}^{p-1} \e({\bf x}_{\ell}{\bf x}_{\ell+\tau}^T)e^{-i2\pi k \ell/p}$
is the $k$-th Fourier series coefficient of the periodic covariance
\begin{equation}
\label{eq:xn cyclostat}
\e[({\bf x}_{\ell} - \boldsymbol{\mu}_x)({\bf x}_{\ell+\tau} - \boldsymbol{\mu}_x)^T]
=\sum_{k=0}^{p-1} {\bf R}_{x}^{(\frac{k}{p})}(\tau) e^{i2\pi k \ell/p},
\end{equation}
evaluated at the harmonic cyclic frequencies $(\frac{k}{p})_{k=0,1,...,p-1}$. Specifically, the cyclic autocovariance is given by:
${\bf R}_{x}^{(k/p)}(\tau) \pardef \frac{1}{p}\sum_{\ell=0}^{p-1} \e[({\bf x}_{\ell} - \boldsymbol{\mu}_x(\ell))({\bf x}_{\ell+\tau} - \boldsymbol{\mu}_x(\ell+\tau))^T] e^{-i2\pi k \ell/p}.$

Furthermore, we assume that the spectral matrices
${\bf R}_{x'}(f)$ and ${\bf R}_{x}^{(\frac{k}{p})}(f)$
depend on a parameter vector $\boldsymbol{\theta}=(\theta_1,..,\theta_q)^T \in \mathbb{R}^q$. For notational simplicity, this dependence is omitted in the following. Finally, the mappings
$\boldsymbol{\theta} \mapsto {\bf R}_{x'}(f)$ and
$\boldsymbol{\theta} \mapsto {\bf R}_{x}^{(\frac{k}{p})}(f)$
are assumed to be differentiable in a  neighborhood of  the true value of $\boldsymbol{\theta}$, and that the regularity conditions stated in \cite{Delmas2026}
hold.
%
\subsection{Asymptotic FIM rate for cyclostationary CGP with cycle period $p$}
\label{sec:Asymptotic FIM rate for cyclostationary CGP with cycle period p}
%
Since a cyclostationary CGP $({\bf x}_k)_{k\in\mathbb Z}$ with cycle
period $p$ gives rise to the stationary CGP
$({\bf x}'_k)_{k\in\mathbb Z}$ defined above, the asymptotic
frequency-domain FIM rate in \eqref{eq:Whittle's formula} can be
applied to ${\bf x}'_k$. However, because the  blocked process consists of only $q = n/p$ vector samples, the FIM rate for the original process scales by a factor of $1/p$. By substituting the parameters of the blocked process and accounting for this scaling, the terms $(\frac{1}{2}, c)$, ${\bf R}_{x}(f)$, $\boldsymbol{\mu}_{x,k}'$, and ${\bf R}_{x,k}'(f)$ are replaced by $(\frac{1}{2p}, \frac{c}{p^2})$, $\boldsymbol{\mu}_{x',k}'$, ${\bf R}_{x'}(f)$, and ${\bf R}_{x',k}'(f)$, respectively.

For a cyclostationary CGP, it is computationally and physically more insightful to express the resulting asymptotic frequency-domain FIM rate directly in terms of the cyclic spectra $\mathbf{R}_{x}^{(k/p)}(f)$ for $k=0,\dots,p-1$, rather than relying on the conventional spectral matrix $\mathbf{R}_{x'}(f)$ of the  blocked process.
Building upon the foundational work of Gladyshev \cite{Gladyshev1961} on periodically correlated  random sequences (see also \cite[eq. (2.7)]{Nematollahi2005} and \cite[Proposition 6.9]{Hurd2007}), this concept extends naturally to multidimensional cyclostationary processes. Specifically, the conventional spectral matrix ${\bf R}_{x'}(f)$ is unitarily similar to an $mp \times mp$ symmetric positive definite poly-spectral matrix, denoted ${\bf R}_{x}^{\rm cyc}(f)$.  Its $(k',l')$-th $m \times m$ block of ${\bf R}_{x}^{\rm cyc}(f)$ is defined as function of the cyclic spectra ${\bf R}_{x}^{(\frac{k}{p})}(f)$ by the following matrices  \cite[eq. (2.2)]{Nematollahi2005}:
\begin{equation}
\label{eq:cyclic spectral matrix}
\frac{1}{p}{\bf R}_{x}^{(\frac{\ell'-k'}{p})}\left(\frac{f-k'+1}{p}\right),\ k',\ell'=1,...,p.
\end{equation}
This relationship yields the following block EVD  of ${\bf R}_{x'}(f)$:
\begin{equation}
\label{eq:unitarily similar}
{\bf R}_{x'}(f)
= p
[{\bf U}_p(f) \otimes {\bf I}_m]^H{\bf R}_{x}^{\rm cyc}(f)[{\bf U}_p(f) \otimes {\bf I}_m],
\end{equation}
where ${\bf U}_p(f)$ is the $p \times p$ unitary  DFT matrix with entries defined by $[{\bf U}_p(f)]_{k',\ell'}\pardef \frac{1}{\sqrt{p}}e^{i2\pi(k'-1-f)(\ell'-1)/p}$. 
Substitution of the block EVD \eqref{eq:unitarily similar} into \eqref{eq:Whittle's formula} for the blocked process, and exploiting the unitary invariance of the trace operator, the factors of $p$ inside the trace operations cancel out elegantly. This leads to the following fundamental result:
\begin{result}
\label{result:cyclicWhittle formula}
For a multidimensional cyclostationary real-valued CGP with cycle period $p$, the asymptotic frequency-domain FIM rate is given by:
\begin{eqnarray}
\nonumber
\lim_{n \rightarrow \infty}
\frac{1}{n}\left({\rm FIM}_{{\bf y}_n}(\boldsymbol{\theta})\right)_{k,\ell}
&=&
\frac{1}{p}\e\left(\tau^{-1}\right) ({\boldsymbol \mu}^{{\rm cyc}^{'}}_{x,k})^T
[{\bf R}_{x}^{{\rm cyc}}(0)]^{-1}
{\boldsymbol \mu}_{x,\ell}^{{\rm cyc}^{'}}
\\
\nonumber
&+&
\frac{1}{2p}\int_{0}^{1}\tra([{\bf R}_{x}^{{\rm cyc}}(f)]^{-1} {\bf R}_{x,k}^{{\rm cyc}^{'}}(f)[ {\bf R}_{x}^{{\rm cyc}}(f)]^{-1}  {\bf R}_{x,\ell}^{{\rm cyc}^{'}}(f))df
\\
\label{eq:cyclic Whittle's formula}
&+&
\frac{c}{p^2}
\int_{0}^{1}\tra([{\bf R}_{x}^{{\rm cyc}}(f)]^{-1} {\bf R}_{x,k}^{{\rm cyc}^{'}}(f))df
\int_{0}^{1}\tra([{\bf R}_{x}^{{\rm cyc}}(f)]^{-1} {\bf R}_{x,\ell}^{{\rm cyc}^{'}}(f))df,
\end{eqnarray}
where ${\bf R}_{x}^{\rm cyc}(0) \pardef {\bf R}_{x}^{\rm cyc}(f)\big|_{f=0}$, the derivatives are defined as ${\bf R}_{x,k}^{\mathrm{cyc}\prime}(f) \pardef\frac{\partial {\bf R}_{x}^{\rm cyc}(f)}{\partial \theta_k}$, and the transformed mean derivative vector is $\boldsymbol{\mu}_{x,k}^{\mathrm{cyc}\prime} \pardef({\bf W}_p \otimes {\bf I}_m)\boldsymbol{\mu}_{x',k}'$. Here, ${\boldsymbol \mu}_{x',k}' = [\boldsymbol{\mu}_{x_1,k}^{\prime T}, \dots, \boldsymbol{\mu}_{x_p,k}^{\prime T}]^T$, and ${\bf W}_p \in \mathbb{C}^{p \times p}$ is the standard DFT matrix with entries $[{\bf W}_p]_{k,l}\pardef \frac{1}{\sqrt{p}}e^{i2\pi (k-1)(l-1)/p}$.
\end{result}
%
\section{Asymptotic FIM rate for stationary, cyclostationary and almost stationary complex-valued CGP}
\label{sec:Asymptotic FIM rate for stationary, cyclostationary and almost stationary complex-valued CGP}
%
In numerous engineering applications, multidimensional complex-valued processes, whether circular or noncircular, arise naturally.
In these instances, A complex-valued process $({\bf x}_k)_{k \in \mathbb{Z}}$ with ${\bf x}_k \in \mathbb{C}^m$ is defined as a complex CGP if its associated real-valued process $(\bar{\bf x}_k)_{k \in \mathbb{Z}}$ with $\bar{\bf x}_k \pardef [\mathrm{Re}({\bf x}_k)^T, \mathrm{Im}({\bf x}_k)^T]^T \in \mathbb{R}^{2m}$ is a real CGP.
The one-to-one mapping $\bar{\bf x} \mapsto \widetilde{\bf x}\pardef ({\bf x}^T,{\bf x}^H)^T= \sqrt{2}{\bf M}\bar{\bf x}$, where
${\bf M}\pardef
\tiny{\frac{1}{\sqrt{2}}\left(\begin{array}{cc}
{\bf I}& i{\bf I}\\
{\bf I} &  -i{\bf I}\\
\end{array}
\right)}$ is unitary, is subsequently referred  to as a real-complex representation.  It provides a convenient
framework for extending the asymptotic frequency-domain FIM rates in \eqref{eq:Whittle's formula} and \eqref{eq:cyclic Whittle's formula} to both circular and
noncircular complex-valued stationary and cyclostationary CGPs.
%
\subsection{Asymptotic FIM rate for stationary and cyclostationary complex-valued CGP}
\label{sec:Asymptotic FIM rate for stationary and cyclostationary complex-valued CGP}
%
\subsubsection{Complex-valued extension of \eqref{eq:Whittle's formula}}
The real-to-complex representation of CGP is employed, and \eqref{eq:Whittle's formula} is applied to  the associated real-valued process  $(\bar{\bf x}_k)_{k \in \mathbb{Z}} \in \mathbb{R}^{2m}$. This yields a direct extension of \eqref{eq:Whittle's formula} to stationary complex-valued CGP, with the corresponding parameters and functions summarized in Table 1. In particular,
${\bf C}_{\tilde{x}}(f)
=\tiny{\left(\begin{array}{cc}
{\bf C}_{x}(f)& {\bf R}_{x}(f)\\
{\bf R}_{x}^*(-f) &  {\bf C}^*_{x}(-f)\\
\end{array}
\right)}
=\sum_{k\in\mathbb{Z}}{\bf C}_{\tilde{x}}(k)e^{-i2\pi k f}$, where the block matrices ${\bf C}_{x}(f)\pardef \sum_{k\in\mathbb{Z}}{\bf C}_{x}(k)e^{-i2\pi k f}$ and
 ${\bf R}_{x}(f)\pardef\sum_{k\in\mathbb{Z}}{\bf R}_{x}(k)e^{-i2\pi k f}$ represent the spectrum  and  pseudo-spectrum matrices, respectively,
and where
${\bf C}_{\tilde{x}}(k) \pardef \e[(\tilde{\bf x}_{\ell} - \tilde{\boldsymbol{\mu}}_x)(\tilde{\bf x}_{\ell+k} - \tilde{\boldsymbol{\mu}}_x)^H]$ with 
$\tilde{\boldsymbol{\mu}}_x \pardef ({\boldsymbol{\mu}}_x^T,{\boldsymbol{\mu}}_x^H)^T$, ${\bf C}_{x}(k) \pardef \e[({\bf x}_{\ell} - \boldsymbol{\mu}_x)({\bf x}_{\ell+k} - \boldsymbol{\mu}_x)^H]$ and 
${\bf R}_{x}(k) \pardef \e[({\bf x}_{\ell} - \boldsymbol{\mu}_x)({\bf x}_{\ell+k} - \boldsymbol{\mu}_x)^T]$. 
%
\begin{center}
\scriptsize{
\begin{tabular}{|c|c|c|c|c|c|}
\hline
real & real & noncircular complex & circular complex& noncircular complex & circular complex\\
stationary & cyclostationary & stationary & stationary & cyclostationary & cylostationary\\
\hline
$(\frac{1}{2},c)$&$(\frac{1}{2p},\frac{c}{p^2})$&$(\frac{1}{2},c)$&$(1,2c)$&$(\frac{1}{2p},\frac{c}{p^2})$&$(\frac{1}{p},\frac{2c}{p^2})$\\
$\!\!{\boldsymbol \mu}^{'T}_{x,k}{\bf R}_{x}^{-1}(0){\boldsymbol \mu}'_{x,\ell}$\!\!&\!\!${\boldsymbol \mu}^{'T}_{x',k}{\bf R}_{x'}^{-1}(0){\boldsymbol \mu}'_{x',\ell}$\!\!&\!\!\!${\boldsymbol \mu}^{'H}_{\tilde{x},k}{\bf C}_{\tilde{x}}^{-1}(0){\boldsymbol \mu}'_{\tilde{x},\ell}\!\!\!\!$&\!\!$2{\rm Re}\{{\boldsymbol \mu}^{'H}_{x,k}{\bf C}_{x}^{-1}(0){\boldsymbol \mu}'_{x,\ell}\}$\!\!
&\!\!$\tilde{\boldsymbol \mu}_{x',k}{\bf C}_{\tilde{x}'}^{-1}(0)\tilde{\boldsymbol \mu}_{x',\ell}$\!\!&
\!\!$2{\rm Re}\{{\boldsymbol \mu}^{'H}_{x',k}{\bf C}_{x'}^{-1}(0){\boldsymbol \mu}'_{x',\ell}\}$\!\!\\
\!\!\!\!${\bf R}_{x}(f)$, ${\bf R}'_{x,k}(f)$\!\!\!&\!\!\!${\bf R}_{x'}(f)$, ${\bf R}'_{x',k}(f)$\!\!\!&${\bf C}_{\tilde{x}}(f)$, ${\bf C}'_{\tilde{x},k}(f)$&
${\bf C}_x(f)$, ${\bf C}'_{x,k}(f)$&${\bf C}_{\tilde{x}'}(f)$, ${\bf C}'_{\tilde{x}',k}(f)$&\!\!\!${\bf C}_{x'}(f)$, ${\bf C}'_{x',k}(f)$\!\!\!\\
\hline
\end{tabular}
}
\end{center}
\centerline{
{\footnotesize
Table.1. Parameters and functions involved in the asymptotic FIM rate for real and complex-valued CGP.}
}
%
\subsubsection{Complex-valued extension of Result \ref{result:cyclicWhittle formula}}
The noncircular complex-valued cyclostationary CGP with cycle period $p$
are also characterized by their cyclic spectra, which are expressed as:
${\bf C}_{\tilde{x}}^{(\frac{k}{p})}(f)\pardef \sum_{\tau \in\mathbb{Z}}{\bf C}_{\tilde{x}}^{(\frac{k}{p})}(\tau)e^{-i2\pi \tau f}$, $k=0,1,..,p-1$,
 where 
${\bf C}_{\tilde{x}}^{(\frac{k}{p})}(\tau)
\pardef
\frac{1}{p}\sum_{\ell=0}^{p-1} \e(\widetilde{\bf x}_{\ell} - \tilde{\boldsymbol{\mu}}_x)(\widetilde{\bf x}_{\ell+\tau} - \tilde{\boldsymbol{\mu}}_x)^H] )e^{-i2\pi k \ell/p}$
is the $k$-th Fourier series coefficient of the periodic augmented covariance:
$\e[(\widetilde{\bf x}_{\ell} - \tilde{\boldsymbol{\mu}}_x)(\widetilde{\bf x}_{\ell+\tau} - \tilde{\boldsymbol{\mu}}_x)^H] 
=\sum_{k=0}^{p-1} {\bf C}_{\tilde{x}}^{(\frac{k}{p})}(\tau) e^{i2\pi k \ell/p}$ evaluated  at the harmonic cyclic frequencies $(\frac{k}{p})_{k=0,1,...,p-1}$. 

Using the one-to-one mapping $\bar{\bf x} \mapsto \widetilde{\bf x}= \sqrt{2}{\bf M}\bar{\bf x}$
and the $2mp \times 2mp$ unitary matrix ${\rm Blkdiag}({\bf M},...,{\bf M})[{\bf U}_p(f)\otimes {\bf I}_{2m}]$, we straightforwardly obtain the following result from Result \ref{result:cyclicWhittle formula}:
\begin{result}
\label{result:complex cyclicWhittle formula}
The asymptotic frequency-domain FIM rate for multidimensional cyclostationary noncircular complex-valued CGP is given by:
\begin{eqnarray}
\nonumber
\lim_{n \rightarrow \infty}
\frac{1}{n}\left({\rm FIM}_{{\bf y}_n}(\boldsymbol{\theta})\right)_{k,\ell}
&=&
\frac{1}{p}\e\left(\tau^{-1}\right)
 (\widetilde{{\boldsymbol \mu}^{{\rm cyc}^{'}}_{x,k}})^H
[{\bf C}_{\tilde{x}}^{{\rm cyc}}(0)]^{-1}
\widetilde{{\boldsymbol \mu}_{x,\ell}^{{\rm cyc}^{'}}}
\\
\nonumber
&+&
\frac{1}{2p}\int_{0}^{1}\tra([{\bf C}_{\tilde{x}}^{{\rm cyc}}(f)]^{-1} {\bf C}_{\tilde{x},k}^{{\rm cyc}^{'}}(f) [{\bf C}_{\tilde{x}}^{{\rm cyc}}(f)]^{-1}
{\bf C}_{\tilde{x},\ell}^{{\rm cyc}^{'}}(f))df
\\
\label{eq:complex cyclic Whittle's formula}
&+&
\frac{c}{p^2}
\int_{0}^{1}\tra([{\bf C}_{\tilde{x}}^{{\rm cyc}}(f)]^{-1} {\bf C}_{\tilde{x},k}^{{\rm cyc}^{'}}(f))df
\int_{0}^{1}\tra([{\bf C}_{\tilde{x}}^{{\rm cyc}}(f)]^{-1} {\bf C}_{\tilde{x},\ell}^{{\rm cyc}^{'}}(f))df,
\end{eqnarray}
where 
$\widetilde{{\boldsymbol \mu}^{{\rm cyc}^{'}}_{x,k}}\pardef (({\boldsymbol \mu}^{{\rm cyc}^{'}}_{x,k})^T,({\boldsymbol \mu}^{{\rm cyc}^{'}}_{x,k})^H)^T$,
${\bf C}_{\tilde{x}}^{{\rm cyc}}(0)\pardef {\bf C}_{\tilde{x}}^{{\rm cyc}}(f)_{|f=0}$
and ${\bf C}_{\tilde{x},k}^{\mathrm{cyc}\prime}(f) \pardef\frac{\partial {\bf C}_{\tilde{x}}^{\rm cyc}(f)}{\partial \theta_k}$.
The $(k',\ell')$th $2m \times 2m$ block of the augmented poly-spectrum matrix ${\bf C}_{\tilde{x}}^{\rm cyc}(f)$ is defined by
$\frac{1}{p}{\bf C}_{\tilde{x}}^{(\frac{\ell'-k'}{p})}\left(\frac{f-k'+1}{p}\right),\ k',\ell'=1,...,p.$ where
${\bf C}_{\tilde{x}}^{(\frac{k}{p})}(f)
\pardef
\tiny{\left(\begin{array}{cc}
{\bf C}^{(\frac{k}{p})}_{x}(f)& {\bf R}^{(\frac{k}{p})}_{x}(f)\\
{\bf R}^{(-\frac{k}{p})^*}_{x}(-f) &  {\bf C}^{(-\frac{k}{p})^*}_{x}(-f)\\
\end{array}
\right)}$.

For multidimensional cyclostationary circular complex-valued CGPs, the pseudo-spectral matrices vanish (i.e., ${\bf R}_{x}^{(k/p)}(f) = {\bf 0}$), making ${\bf C}_{\tilde{x}}^{\rm cyc}(f)$ block-diagonal. Consequently, \eqref{eq:complex cyclic Whittle's formula} simplifies and can be expressed directly in terms of the $m \times m$ spectral matrices ${\bf C}_{x}^{\rm cyc}(f)$ by replacing
$(1/2p,c/p^2), (\widetilde{{\boldsymbol \mu}^{{\rm cyc}^{'}}_{x,k}})^H
[{\bf C}_{\tilde{x}}^{{\rm cyc}}(0)]^{-1}
\widetilde{{\boldsymbol \mu}_{x,\ell}^{{\rm cyc}^{'}}}$,
${\bf C}_{\tilde{x}}^{{\rm cyc}}(f)$ and ${\bf C}_{\tilde{x},k}^{{\rm cyc}^{'}}(f)$
by
$(1/p,2c/p^2)$, $2{\rm Re}[({\boldsymbol \mu}^{{\rm cyc}^{'}}_{x,k})^H
[{\bf C}_{x}^{{\rm cyc}}(0)]^{-1}
{\boldsymbol \mu}_{x,\ell}^{{\rm cyc}^{'}}]$, ${\bf C}_{x}^{{\rm cyc}}(f)$ and ${\bf C}_{x,k}^{{\rm cyc'}}(f)$, respectively, 
where
the $(k',\ell')$th $m \times m$ block of the poly-spectrum ${\bf C}_{x}^{\rm cyc}(f)$ is defined by
$\frac{1}{p}{\bf C}_{x}^{(\frac{\ell'-k'}{p})}\left(\frac{f-k'+1}{p}\right),\ k',\ell'=1,...,p$, where
 ${\bf C}_{x}^{(\frac{k}{p})}(f)\pardef \sum_{\tau \in\mathbb{Z}}{\bf C}_{x}^{(\frac{k}{p})}(\tau)e^{-i2\pi \tau f}$, $k=0,1,..,p-1$.
\end{result}

Finally, it is important to note that the conjecture proposed in \cite[rels. (6), (7)]{Schell1992} for circular complex Gaussian data, when expressed using our notation, takes the form:
$
\lim_{n \to \infty} \frac{1}{n} \left(\mathrm{FIM}_{{\bf y}_n}(\boldsymbol{\theta})\right)_{k,\ell} \approx \frac{1}{2}\mathrm{tr}\left({\bf S}^{-1}\frac{\partial{\bf S}}{\partial \theta_k}{\bf S}^{-1}\frac{\partial{\bf S}}{\partial \theta_{\ell}}\right),
$
where the $(k',\ell')$-th block of ${\bf S}$ is defined as  $[{\bf S}]_{k',\ell'}\pardef {\bf C}^{(\frac{k'-\ell'}{p})}_x(\frac{k'+\ell'}{2p})$ where ${\bf C}^{(\frac{k}{p})}_x(f)$ represents the  cyclic spectrum. This latter formula is grounded under any proof and differs significantly from Result \ref{result:complex cyclicWhittle formula}  based on rigorous reasoning.
%
%
\subsection{Application to linearly modulated digital communication signals}
\label{sec:Application to linearly modulated digital communication signals}
%
We give in this subsection meaningful interpretations of the various spectral matrices introduced
previously, namely,  ${\bf C}_{\tilde{x}'}(f)$, ${\bf C}_{x'}(f)$
${\bf C}_{\tilde{x}}^{\rm cyc}(f)$ and
${\bf C}_{x}^{\rm cyc}(f)$, by applying them to the case of 
noisy over-sampled pulse amplitude modulated signals. The continuous-time signal model is given by
${\bf x}(t) = \sum_{\ell \in\mathbb{Z}} b_{\ell}{\bf h}(t-\ell T)+{\bf w}(t) \in  \mathbb{C}^m,  t \in  \mathbb{R}$ 
where $(b_{\ell})_{\ell \in \mathbb{Z}} \in \mathbb{C}$ is a sequence of zero-mean independent identically distributed symbols with variance
$\sigma_b^2\pardef \e|b_{\ell}|^2$, 
${\bf h}(t)\in \mathbb{C}^m$ is the overall impulse response and ${\bf w}(t)$ is an additive, circular,  temporally and spatially white noise with spectra $C_w(f)=\sigma_w^2{\bf I}$.
The over-sampled discrete-time process, denoted as ${\bf x}_k \pardef {\bf x}(t)_{|t=kT/p}$,  is cyclostationary with a cycle period equal to the oversampling rate $p$, and  reduces to a stationary process when $p=1$. Depending on whether the symbols $b_{\ell}$ are real- or complex-valued, the process
is noncircular or circular, respectively.

For this signal model, the conventional spectra of the blocked augmented process  $\widetilde{\bf x}'_k$ and the blocked process ${\bf x}'_k$ are respectively  given by
${\bf C}_{\tilde{x}'}(f)=\sigma_b^2\widetilde{\bf h}'(f)\widetilde{\bf h}^{'H}(f)+\sigma_w^2{\bf I}$
and
${\bf C}_{x'}(f)=\sigma_b^2{\bf h}'(f){\bf h}^{'H}(f)+\sigma_w^2{\bf I}$
where $\widetilde{\bf h}'(f)\pardef\sum_{k\in\mathbb{Z}}\widetilde{\bf h}'_ke^{-i2\pi kf}$
and
${\bf h}'(f)\pardef \sum_{k \in\mathbb{Z}}{\bf h}'_ke^{-i2\pi kf}$. The blocked impulse response vectors are defined as $\widetilde{\bf h}'_k \pardef [\widetilde{\bf h}^T(t)|_{t=kT+T/p}, \dots, \widetilde{\bf h}^T(t)|_{t=kT+pT/p}]^T$ with $\widetilde{\bf h}(t) \pardef [{\bf h}^T(t), {\bf h}^H(t)]^T$, and ${\bf h}'_k \pardef [{\bf h}^T(t)|_{t=kT+T/p}, \dots, {\bf h}^T(t)|_{t=kT+pT/p}]^T$.

Under the simplifying assumption that $\widetilde{\bf x}_k$ and ${\bf x}_k$ are zero-mean cyclostationary CGP, the asymptotic
frequency-domain FIM rate extensions summarized in the last two columns of Table 1 apply directly. When
applying these results, it is assumed that the channel responses $\widetilde{\bf h}'(f)$ and ${\bf h}'(f)$ are known up to a finite number
of unknown parameters (e.g., in specular multipath channels). It is worth noting that this formulation differs from standard
blind channel estimation models for which (a)  the number of consecutive non-zero time taps in $\widetilde{\bf h}'_k$ and ${\bf h}'_k$ is finite and determined by the channel order; (b) the symbols $b_{\ell}$ are deterministic, unknown parameters; and (c) specific constraints are imposed on the time taps to ensure  identifiability (see, e.g., \cite{Stoica2001} and references therein).

The associated cyclic spectra are given by
${\bf C}_{\tilde{x}}^{(\frac{k}{p})}(f)
=\frac{\sigma_b^2}{p}\widetilde{\bf h}(f)\widetilde{\bf h}^H(f-\frac{k}{p})+\delta_{k,0}\sigma_w^2{\bf I}$
and
${\bf C}_{x}^{(\frac{k}{p})}(f)=\frac{\sigma_b^2}{p}{\bf h}(f){\bf h}^H(f-\frac{k}{p})+\delta_{k,0}\sigma_w^2{\bf I}$ for $k=0,\ldots,p-1$, where 
$\widetilde{\bf h}(f)\pardef \sum_k\widetilde{\bf h}_ke^{-i2\pi kf}$
and ${\bf h}(f)\pardef \sum_k {\bf h}_ke^{-i2\pi kf}$, with $\widetilde{\bf h}_k \pardef ({\bf h}^T_k,{\bf h}^H_k)^T$, and
 ${\bf h}_k \pardef {\bf h}(t)_{|t=kT/p}$ (see, e.g., \cite[subsect. 17.4.3]{Giannakis2010} for equivalent expressions with different notation). 
Consequently, it is straightforward to prove that the poly-spectral matrices
defined in Result \ref{result:complex cyclicWhittle formula} exhibit the following rank-one structural characteristics: 
\begin{equation}
\label{eq:spectre cyclique}
{\bf C}_{\tilde{x}}^{\rm cyc}(f)=\frac{\sigma_b^2}{p^2}\widetilde{\bf h}^{\rm cyc}(f)(\widetilde{\bf h}^{\rm cyc}(f))^H+ \frac{\sigma_w^2}{p}{\bf I}
\ \mbox{and}\
 {\bf C}_{x}^{\rm cyc}(f)=\frac{\sigma_b^2}{p^2}{\bf h}^{\rm cyc}(f)({\bf h}^{\rm cyc}(f))^H+ \frac{\sigma_w^2}{p}{\bf I}
\end{equation}
where the stacked frequency responses are defined as:
$\widetilde{\bf h}^{\rm cyc}(f)\pardef (\widetilde{\bf h}^T(\frac{f}{p}),\widetilde{\bf h}^T(\frac{f-1}{p}),..., \widetilde{\bf h}^T(\frac{f-p+1}{p}) )^T$
and ${\bf h}^{\rm cyc}(f)\pardef ({\bf h}^T(\frac{f}{p}),{\bf h}^T(\frac{f-1}{p}),..., {\bf h}^T(\frac{f-p+1}{p}) )^T$.
The expressions in \eqref{eq:spectre cyclique} for the poly-spectrum, which were introduced by Gladyshev in \cite{Gladyshev1961}, provide a meaningful interpretation and, to the best of our knowledge, represent a novel contribution to the field.

Finally, it is important to that over-sampled  offset QAM and, more generally, over-sampled quasi-rectilinear modulations \cite{Chevalier2018}
${\bf x}(t)=\sum_{\ell}i^{\ell} b_{\ell}{\bf h}(t-\ell T)$,
where $b_{\ell}$ are real-valued, are also noncircular cyclostationary but,
with a period equal to twice the oversampling rate $p$ because
$\e({\bf x}_{k+ jp}{\bf x}_{\ell+ jp}^H)=\e({\bf x}_{k}{\bf x}_{\ell}^H)$ and
$\e({\bf x}_{k+ j2p}{\bf x}_{\ell+ j2p}^T)=\e({\bf x}_{k}{\bf x}_{\ell}^T)$
for arbitrary $ j\in \mathbb{Z}$.
%
\subsection{Extension to particular multidimensional (wide-sense) almost cyclostationary complex-valued CGP}
\label{sec:Extension to particular multidimensional almost cyclostationary complex-valued CGP}
%
We notice that the methodology developed in this paper does not, in its present form, cover a general
zero-mean wide-sense almost cyclostationary CGP, for which the covariance and pseudo-covariance sequences
may contain arbitrary, not necessarily harmonically related, cycle frequencies. More precisely, a zero-mean
complex-valued process is wide-sense almost cyclostationary if there exist sets of cycle frequencies $(\alpha_k)_{k\in\mathbb{Z}}$ and $(\beta_k)_{k\in\mathbb{Z}}$ such that\footnote{
Comparing \eqref{eq:def almost cyclostationary} with
\eqref{eq:xn cyclostat}, we see that wide-sense cyclostationarity is a
special case of wide-sense almost cyclostationarity in which the cycle
frequencies are restricted to integer multiples of the fundamental cycle
frequency $1/p$.}
\begin{equation}
\label{eq:def almost cyclostationary}
\e\!\left({\bf x}_{\ell}{\bf x}_{\ell+\tau}^{T}\right)
=
\sum_{k\in\mathbb{Z}}
{\bf R}_{x}^{\alpha_k}(\tau)
e^{i2\pi\alpha_k\ell},
\qquad
\e\!\left({\bf x}_{\ell}{\bf x}_{\ell+\tau}^{H}\right)
=
\sum_{k\in\mathbb{Z}}
{\bf C}_{x}^{\beta_k}(\tau)
e^{i2\pi\beta_k\ell}.
\end{equation}

Although a general almost cyclostationary process is beyond the scope of
the present methodology, an important practically relevant subclass can
be treated. Consider
\begin{equation}
\label{eq:almost cyclostationary}
{\bf x}_k
\pardef
e^{i2\pi k\theta_0}{\bf u}_k
\in\mathbb{C}^m,
\qquad k\in\mathbb{Z},
\end{equation}
where $({\bf u}_k)_{k\in\mathbb{Z}}$ is a zero-mean circular or
noncircular complex-valued CGP that is either stationary or
cyclostationary with period $p$. This model describes a common impairment
in digital communications, namely, a carrier-frequency offset (CFO)
$\theta_0$ resulting from a frequency mismatch between the transmitter
and receiver local oscillators.

The effect of the frequency offset on the second-order statistics depends
on the circularity and cyclostationarity of the underlying process. In
particular, if ${\bf u}_k$ is circular and stationary, then ${\bf x}_k$
remains circular and stationary, with spectral density matrix
${\bf C}_x(f)={\bf C}_u(f+\theta_0)$.
The other cases are summarized in Table 2.
\begin{center}
\scriptsize{
\begin{tabular}{|c|c|}
\hline
${\bf u}_k$ & ${\bf x}_k=e^{i2\pi k\theta_0}{\bf u}_k$ \\
\hline
circular stationary&circular stationary\\
\hline
noncircular stationary&noncircular cyclostationary with cyclic frequencies $(\alpha_0,\beta_0)=(2\theta_0,0)$\\
\hline
circular cyclostationary with  cycle period $p$&circular cyclostationary with  cycle period $p$\\
\hline
\!\!\!\! noncircular cyclostationary with  cycle period $p$ \!\!\!\!& \!\!noncircular almost cyclostationary with cyclic frequencies 
$(\alpha_k,\beta_k)\!=\!(\frac{k}{p}\!+\!2\theta_0,\frac{k}{p})$, $k\!=\!0,...,p\!-\!1$\!\!\\
\hline
\end{tabular}
}
\end{center}
\centerline{
{\footnotesize
Table.2. Properties of $({\bf u}_k,{\bf x}_k$) under a frequency offset.}
}

When $({\bf u}_{k})_{k\in \mathbb{Z}}$ is an arbitrary (non necessarily stationary or cyclostationary) zero-mean CGP, the r.v. ${\bf y}_n=({\bf x}_{1}^T,...,{\bf x}_{n}^T)^T  \in \mathbb{C}^{mn}$ where ${\bf x}_{k}$ is given by \eqref{eq:almost cyclostationary}, is therefore also CG distributed.
Applying the complex-valued extension of the Slepian-Bangs formula for
CG distributions established in \cite{Abeida2019} to ${\bf y}_n$, yields the following
result proven in the Appendix.
\begin{result}
\label{result:almost cyclo Whittle formula}
For arbitrary circular [resp. noncircular] zero-mean complex-valued CGP
$({\bf u}_k)_{k \in \mathbb{Z}}$ affected by the frequency offset $\theta_0$, the parameter $\theta_0$
and the parameters $(\theta_1,....,\theta_q)$ of the
covariance
$\boldsymbol{\Sigma}_{{\bf z}_n}\pardef \e({\bf z}_n{\bf z}_n^H)$ where ${\bf z}_n \pardef ({\bf u}_{1}^T,...,{\bf u}_{n}^T)^T$
[resp.
$\boldsymbol{\Sigma}_{\tilde{\bf z}_n}\pardef
\e(\widetilde{\bf z}_n\widetilde{\bf z}_n^H)$, where $\widetilde{\bf z}_n \pardef (\widetilde{\bf u}_{1}^T,...,\widetilde{\bf u}_{n}^T)^T$],
are decoupled in the $(q+1) \times (q+1)$
${\rm FIM}_{{\bf y}_n}(\boldsymbol{\theta})$ (i.e., $[{\rm FIM}_{{\bf y}_n}(\boldsymbol{\theta})]_{0,k}= 0$ for $k=1,..,q$) and furthermore
$[{\rm FIM}_{{\bf y}_n}(\boldsymbol{\theta})]_{k,\ell}=[{\rm FIM}_{{\bf z}_n}(\boldsymbol{\theta})]_{k,\ell}$ for
$k,\ell=1,..,q$.
\end{result}

Consequently, whenever the underlying process ${\bf u}_k$ satisfies the
assumptions of the stationary or cyclostationary frequency-domain results
developed above, the corresponding asymptotic frequency-domain FIM rate for the covariance
parameters $(\theta_1,\ldots,\theta_q)$ is obtained directly by applying
the complex stationary formula \eqref{eq:Whittle's formula} (as summarized in Table 1) and
Result~\ref{result:complex cyclicWhittle formula} to ${\bf u}_k$, without
explicitly incorporating the frequency offset $\theta_0$. Thus, for this
single-component model, the frequency offset does not modify the
asymptotic frequency-domain FIM rate for the parameters governing the covariance and
pseudo-covariance spectra of the underlying process.

Section~\ref{sec:Application to DOA estimation} provides an illustrative
application of this result, including a closed-form expression for the
asymptotic CRB on  $\theta_0$ for the model
\eqref{eq:almost cyclostationary}.
It is also instructive to consider the multi-component model
${\bf x}_k=\sum_{\ell=1}^{q_0}e^{i2\pi k\theta_{0,\ell}}{\bf u}_{k,\ell}$,
where the processes $({\bf u}_{k,\ell})_{\ell=1,\ldots,q_0}$ are mutually
uncorrelated, circular, and stationary, with spectra
${\bf C}_{u_\ell}(f)$. In this case, ${\bf x}_k$ is circular and
stationary and has spectrum
${\bf C}_x(f)=\sum_{\ell=1}^{q_0}{\bf C}_{u_\ell}(f+\theta_{0,\ell})$.
Therefore, the complex stationary extension of
\eqref{eq:Whittle's formula} applies directly, with parameter vector
$\boldsymbol{\theta}=(\theta_{0,1},\ldots,\theta_{0,q_0},\theta_{1,1},\ldots,\theta_{1,q_1},\ldots,\theta_{q_0,1},\ldots,\theta_{q_0,q_{q_0}})$,
where $(\theta_{\ell,1},\ldots,\theta_{\ell,q_\ell})$ parameterizes
${\bf C}_{u_\ell}(f)$.

In contrast to the single-component model, the frequency-offset
parameters $(\theta_{0,1},\ldots,\theta_{0,q_0})$  and the spectral parameters 
$(\theta_{1,1},\ldots,\theta_{1,q_1},\ldots,\theta_{q_0,1},\ldots,\theta_{q_0,q_{q_0}})$ are generally no longer
FIM-decoupled. Consequently, the different frequency offsets
can affect the CRBs of the spectral parameters, and conversely; see,
e.g., \cite{Ladaycia2017}. A subsequent section provides an illustrative
example showing the impact of frequency offsets on the CRB for DOA
estimation.
%
\section{Application to DOA estimation}
\label{sec:Application to DOA estimation}
%
This section illustrates the influence of the spectral characteristics of the source signals
on the CRB for estimating DOA parameters, including scenarios in which the measurements are affected by a carrier
frequency offset. More
specifically, we consider  one or two narrowband far-field signals impinge on an arbitrary array of $m$ sensors.
After frequency down-shifting the filtered sensor signals to baseband, the generated complex envelope, sampled at time $k$ is modeled as
\begin{equation}
\label{eq:DOA model}
{\bf x}_k
={\bf A}{\bf s}_k+ {\bf n}_k \in \mathbb{C}^{m}.
\end{equation}
The steering matrix ${\bf A}$ is either ${\bf A}= {\bf a}_1$ or $[{\bf a}_1,{\bf a}_2]$ where ${\bf a}_i$, $i=1,2$ are the steering vectors
parameterized by the DOA $\omega_i$ with
$\|{\bf a}_i\|^2=m$. We define $\alpha_i \pardef 2{\bf d}_i^H {\bf \Pi}^{\bot}_{{\bf a}_i}{\bf d}_i$ with ${\bf d}_i\pardef \frac{d{\bf a}_i}{d\omega_i}$ and where
${\bf \Pi}^{\bot}_{{\bf a}_i}$ denotes the orthogonal projector onto the orthogonal complement of ${\bf a}_i$.
The measurement noise $({\bf n}_k)_{k \in \mathbb{Z}}$ is is assumed to be zero-mean, circular complex-valued, spatially uncorrelated with $\e({\bf n}_k{\bf n}_k^H)=\sigma_n^2 {\bf I}$ and white stationary with spectrum
${\bf C}_n(f)=\sigma_n^2 {\bf I}$.
The source signals, denoted as ${\bf s}_k=s_{k,1}$ or $(s_{k,1},s_{k,2})^T$ are zero-mean complex-valued, and not correlated with ${\bf n}_k$.
Furthermore, we assume that the process $({\bf x}_k)_{k \in \mathbb{Z}}$ is CG distributed\footnote{In contrast to conventional models, our approach does not require a specific probability distribution for the source and noise signals, thereby encompassing the standard Gaussian assumption as a special case.}.
In contrast to all previously published performance studies,
we assume here that the source signals $({\bf s}_k)_{k \in \mathbb{Z}}$ are not necessarily independent, but are instead stationary, cyclostationary or almost
cyclostationary signals.
%
\subsection{Stationary CGP}
\label{sec:Stationary CGP}
%
In this instance, the source signals $({\bf s}_k)_{k \in \mathbb{Z}}$ are stationary, and are either circular or rectilinear (i.e., strictly second-order noncircular).
The SNRs are defined as $r_i\pardef \frac{\sigma_{s_i}^2}{\sigma_n^2}$ with $\sigma_{s_i}^2\pardef \e(|s_{k,i}|^2)$.

For circular sources with temporal correlation
${\bf C}_s(k)\pardef \e({\bf s}_{\ell}{\bf s}_{\ell+k}^H)$ and spectrum ${\bf C}_s(f)$, model \eqref{eq:DOA model} implies that the observations have covariance
 ${\bf C}_x(k) ={\bf A}{\bf C}_s(k){\bf A}^H+\sigma_n^2 \delta_{k,0}{\bf I}$
and spectrum ${\bf C}_x(f)={\bf A}{\bf C}_s(f){\bf A}^H+\sigma_n^2{\bf I}$.

For rectilinear sources, each source can be written as $s_{k,i}=\rho_{k,i} e^{i\phi_i}$, $i=1,2$ where $\phi_i$ is a fixed (deterministic phase and $\rho_{k,i}\in \mathbb{R}$ is a real-valued amplitude process. Let $\boldsymbol{\rho}_k = \rho_{k,1}$ (single source) or $\boldsymbol{\rho}_k = (\rho_{k,1}, \rho_{k,2})^T$ (two sources), and define the temporal covariance ${\bf R}_\rho(k) \pardef \e(\boldsymbol{\rho}_\ell \boldsymbol{\rho}_{\ell+k}^T)$ with associated spectrum 
${\bf R}_\rho(f)$. 
This results in noncircular measurements ${\bf x}_k$,  which therefore necessitates the use of the extended model of  \eqref{eq:DOA model}, as
\begin{equation}
\label{eq:DOA extended model}
\widetilde{\bf x}_k
=\widetilde{\bf A}{\boldsymbol \rho}_k+ \widetilde{\bf n}_k \in \mathbb{C}^{2m},
\end{equation}
with $\widetilde{\bf A} =  \widetilde{\bf a}_1$ (single source) or $[ \widetilde{\bf a}_1, \widetilde{\bf a}_2]$ (two sources) where $ \widetilde{\bf a}_i \pardef ({\bf a}_i^Te^{i\phi_i},{\bf a}^H_ie^{-i\phi_i})^T$, $i=1,2$.
This model implies that
 ${\bf C}_{\tilde{x}}(k) =\widetilde{\bf A}{\bf R}_{\rho}(k)\widetilde{\bf A}^H+\sigma_n^2 \delta_{k,0}{\bf I}$
 and associated spectrum ${\bf C}_{\tilde{x}}(f) =\widetilde{\bf A}{\bf R}_{\rho}(f)\widetilde{\bf A}^H+\sigma_n^2 {\bf I}$.

In the case of a single source scenario, we assume that the spectrum $C_{s_1}(f)$ (for circular sources) and $R_{\rho_1}(f)$ (for rectilinear sources)
are parameterized by the power $\sigma_{s_1}^2$ and a real-valued parameter\footnote{It is assumed to be  one-dimensional. However, the decoupling properties described in Results \ref{result:DOA une source stat}, \ref{result:DOA 2 sources spectre sans overlap} and \ref{result:DOA une source cyclostationnaire}  apply to a multidimensional parameter as well.}
 $a_1$, such that
$C_{s_1}(f)=\sigma_{s_1}^2 C_{a_1}(f)$ with $\int_0^1 C_{a_1}(f)df=1$ and
$R_{\rho_1}(f)=\sigma_{s_1}^2 R_{a_1}(f)$ with $\int_0^1 R_{a_1}(f)df=1$.
The unknown parameters of our models  are therefore $\boldsymbol{\theta}=(\omega_1,a_1,\sigma_{s_1}^2,\sigma_n^2)^T$ for a single circular source, and 
$\boldsymbol{\theta}=(\omega_1,\phi_1,a_1,\sigma_{s_1}^2,\sigma_n^2)^T$ for a single rectilinear source.
Applying the inversion matrix lemma to the spectra ${\bf C}_x(f)=\sigma_{s_1}^2C_{a_1}(f){\bf a}_1{\bf a}_1^H+\sigma_n^2{\bf I}$
and ${\bf C}_{\tilde{x}}(f)=\sigma_{s_1}^2R_{a_1}(f)\widetilde{\bf a}_1\widetilde{\bf a}_1^H+\sigma_n^2{\bf I}$
 yields the following result, whose main steps of the proof are given in the Appendix.
\begin{result}
\label{result:DOA une source stat}
For a single circular [resp., rectilinear] source, the DOA parameter $\omega_1$ and the parameters $(a_1,\sigma_{s_1}^2,\sigma_n^2)$
[resp., $(\phi_1,a_1,\sigma_{s_1}^2,\sigma_n^2)$] are decoupled in the asymptotic frequency-domain FIM rate \eqref{eq:Whittle's formula}.
Moreover, the DOA-related FIM entries reduce solely on the second term of \eqref{eq:Whittle's formula}. Consequently, the asymptotic CRB for 
 $\omega_1$   is independent of the underlying CG distribution and is given by:
\begin{equation}
\label{eq:Whittle DOA une source circulaire}
\lim_{n \rightarrow \infty}
n{\rm CRB}_c(\omega_1)
=\frac{1}{\alpha_1r_1\int_0^1
\frac{mr_1C_{a_1}^2(f)}{1+mr_1C_{a_1}(f)}df}.
\end{equation}
and
\begin{equation}
\label{eq:Whittle DOA une source rectilinear}
\lim_{n \rightarrow \infty}
n{\rm CRB}_r(\omega_1)
=\frac{1}{2\alpha_1r_1\int_0^1
\frac{mr_1R_{a_1}^2(f)}{1+2mr_1R_{a_1}(f)}df},
\end{equation}
respectively.
Furthermore, these CRBs satisfy the following inequalities:
\begin{equation}
\label{eq:min max c}
\frac{1}{\alpha_1r_1} <
\lim_{n \rightarrow \infty}n{\rm CRB}_c(\omega_1)
\le \frac{1}{\alpha_1r_1}\left(1+\frac{1}{mr_1}\right)
\end{equation}
and
\begin{equation}
\label{eq:min max r}
\frac{1}{\alpha_1r_1} <
\lim_{n \rightarrow \infty}n{\rm CRB}_r(\omega_1)
\le \frac{1}{2\alpha_1r_1}\left(2+\frac{1}{mr_1}\right),
\end{equation}
where the upper bound are achieved when $C_{a_1}(f)=1$ and $R_{a_1}(f)=1$, respectively.
\end{result}

It is worth noting that \eqref{eq:Whittle DOA une source circulaire} and \eqref{eq:Whittle DOA une source rectilinear}
align with the classical expressions previously established for  independent identically complex circular or rectilinear Gaussian signals 
\cite{Delmas2004}, In that case,
$C_{a_1}(f)=1$ and $R_{a_1}(f)=1$, yielding
\begin{equation}
\label{eq:CRB DOA une source gaussien c}
{\rm CRB}_c(\omega_1)
=\frac{1}{n\alpha_1r_1}\left(1+\frac{1}{mr_1}\right)
\end{equation}
and
\begin{equation}
\label{eq:CRB DOA une source gaussien r}
{\rm CRB}_r(\omega_1)
=\frac{1}{2n\alpha_1r_1}\left(2+\frac{1}{mr_1}\right).
\end{equation}

Moreover,  inequalities \eqref{eq:min max c} and  \eqref{eq:min max r} are also consistent with the deterministic CRB (DCRB), obtained when the
source signals are modeled as unknown deterministic parameters. In that case,
${\rm DCRB}_c(\omega_1)
={\rm DCRB}_r(\omega_1)
=\frac{1}{n\alpha_1r_1}$, which is upper bounded by the stochastic CRB.
Furthermore, equalities 
\eqref{eq:Whittle DOA une source circulaire}
and \eqref{eq:Whittle DOA une source rectilinear}
are not overly sensitive to the spectral shape  $C_{a_1}f)$ [resp. $R_{a_1}(f)$]
when $mr_1\gg1$.

We now illustrate Result \ref{result:DOA une source stat} with two examples:

For a flat spectrum\footnote{In the case where  $C_{a_1}(f)=R_{a_1}(f)=\frac{1}{a_1}\mathds{1}_{[b,b+a_1]}(f)$, Fisher's regularity conditions are not satisfied with regard to the parameter $a_1$. Consequently, the unknown parameters of the FIM must be reduced to $(\omega_1,\sigma_{s_1}^2,\sigma_n^2)$.} in $[b,b+a_1] \subset [0,1]$ \eqref{eq:Whittle DOA une source circulaire}, substituting into \eqref{eq:Whittle DOA une source circulaire} and \eqref{eq:Whittle DOA une source rectilinear} yields
\begin{equation}
\label{eq:spectre plat c}
\lim_{n \rightarrow \infty} n\, \mathrm{CRB}_c(\omega_1)
= \frac{1}{\alpha_1 r_1}\left(1 + \frac{a_1}{m r_1}\right),
\end{equation}
and
\begin{equation}
\label{eq:spectre plat r}
\lim_{n \rightarrow \infty} n\, \mathrm{CRB}_r(\omega_1)
= \frac{1}{2\alpha_1 r_1}\left(2 + \frac{a_1}{m r_1}\right).
\end{equation}

For an AR(1) spectrum $C_{a_1}(f) = R_{a_1}(f) = \frac{1-|a_1|^2}{|1-a_1 e^{-i2\pi f}|^2}$ with $|a_1| < 1$. After lengthy algebraic manipulations, \eqref{eq:Whittle DOA une source circulaire} and \eqref{eq:Whittle DOA une source rectilinear} yield
\begin{equation}
\label{eq:spectre AR1 c}
\lim_{n \rightarrow \infty} n\, \mathrm{CRB}_c(\omega_1)
= \frac{1}{\alpha_1 r_1}\left(1 + \frac{1-|a_1|^2}{c-(1-|a_1|^2)}\right),
\end{equation}
and
\begin{equation}
\label{eq:spectre AR1 r}
\lim_{n \rightarrow \infty} n\, \mathrm{CRB}_r(\omega_1)
= \frac{1}{\alpha_1 r_1}\left(1 + \frac{1-|a_1|^2}{c-(1-|a_1|^2)}\right),
\end{equation}
where $c \pardef\left\{(1-|a_1|^2)\left[(1+m_1 r_1)^2 - |a_1|^2(1-m_1 r_1)^2\right]\right\}^{1/2}$,
with $m_1 = m$ for the circular case and $m_1 = 2m$ for the rectilinear case.

It can be verified from \eqref{eq:spectre plat c}--\eqref{eq:spectre plat r} and \eqref{eq:spectre AR1 c}--\eqref{eq:spectre AR1 r} that:
\begin{itemize}
    \item For the flat spectrum \eqref{eq:spectre plat c}--\eqref{eq:spectre plat r}, as $a_1 \to 1$ (white spectrum), the CRBs tend to the Gaussian CRBs \eqref{eq:CRB DOA une source gaussien c}--\eqref{eq:CRB DOA une source gaussien r} (upper bounds). Conversely, as $a_1 \to 0$ (narrowband spectrum), the CRBs tend to the lower bound $\frac{1}{\alpha_1 r_1}$.
    \item For the AR(1) spectrum \eqref{eq:spectre AR1 c}--\eqref{eq:spectre AR1 r}, as $|a_1| \to 0$ (white spectrum), the CRBs tend to the Gaussian CRBs \eqref{eq:CRB DOA une source gaussien c}--\eqref{eq:CRB DOA une source gaussien r} (upper bounds). Conversely, as $|a_1| \to 1$ (narrowband/peak spectrum), the CRBs tend to the lower bound $\frac{1}{\alpha_1 r_1}$.
\end{itemize}

These results confirm that the lower bound $\frac{1}{\alpha_1 r_1}$ in \eqref{eq:min max c}--\eqref{eq:min max r} is approached when the source spectrum becomes highly concentrated (narrowband), while the upper bound is achieved for spectrally white sources.

In the two-source scenario, we assume that the source processes
$(s_{k,1})_{k \in \mathbb{Z}}$
and $(s_{k,2})_{k \in \mathbb{Z}}$ mutually
uncorrelated and are either both circular or both rectilinear. This implies that the spectrum of ${\bf x}_k$ [resp., $\widetilde{\bf x}_k$] is given by
${\bf C}_x(f)=C_{s_1}(f){\bf a}_1{\bf a}_1^H+C_{s_2}(f){\bf a}_2{\bf a}_2^H+\sigma_n^2{\bf I}$
[resp., ${\bf C}_{\tilde{x}}(f)=R_{\rho_1}(f)\widetilde{\bf a}_1\widetilde{\bf a}_1^H+R_{\rho_2}(f)\widetilde{\bf a}_2\widetilde{\bf a}_2^H+\sigma_n^2{\bf I}$]
where the source spectra $C_{s_i}(f)$, [resp., $R_{\rho_i}(f)$] $i=1,2,$  are parameterized by the powers $\sigma_{s_i}^2$ and  real-valued spectral shape parameters $a_i$, such that
$C_{s_i}(f)=\sigma_{s_i}^2 C_{a_i}(f)$ [resp., $R_{\rho_i}(f)=\sigma_{s_i}^2 R_{a_i}(f)$] with $\int_0^1 C_{a_i}(f)df=1$ [resp., $\int_0^1 R_{a_i}(f)df=1$].
The unknown parameter vector is therefore $\boldsymbol{\theta}=(\omega_1,\omega_2,a_1,a_2,\sigma_{s_1}^2,\sigma_{s_2}^2,\sigma_n^2)^T$ for circular sources, and
$\boldsymbol{\theta}=(\omega_1,\omega_2,\phi_1,\phi_2,a_1,a_2,\sigma_{s_1}^2,\sigma_{s_2}^2,\sigma_n^2)^T$ for rectilinear sources.
Applying the matrix inversion lemma to the spectral matrices ${\bf C}_x(f)$ and ${\bf C}_{\tilde{x}}(f)$, as detailed in the
Appendix, yields the following result.
\begin{result}
\label{result:DOA 2 sources spectre sans overlap}
In the circular [resp., rectilinear] two-source model previously described and for spectra without overlap (i.e., spectrum support $B_i\subset[0,1]$, such that $B_1 \cap B_2=\varnothing$),
the DOA parameters $\omega_1$ [resp., $(\omega_1,\phi_1)$] and $\omega_2$ [resp., $(\omega_2,\phi_2)$] are decoupled from each other  in the asymptotic frequency domain FIM rate \eqref{eq:Whittle's formula}.
Moreover, the DOA parameters $(\omega_1,\omega_2)$
[resp., $(\omega_1,\phi_1,\omega_2,\phi_2)$] are FIM-decoupled from $(\sigma_{s_1}^2,\sigma_{s_2}^2,\sigma_n^2)$. 
Consequently, the FIM submatrix associated with the DOA parameters is determined solely by the second term of \eqref{eq:Whittle's formula} and is therefore independent of the CG distribution. The asymptotic CRB for each DOA \(\omega_i\), \(i=1,2\), is then given by
\begin{equation}
\label{eq:Whittle DOA deux sources avec et sans overlap c}
\lim_{n \rightarrow \infty}
n{\rm CRB}_c(\omega_i)
=\frac{1}{\alpha_ir_i\int_{B_i}
\frac{mr_iC_{a_i}^2(f)}{1+mr_iC_{a_i}(f)}df}
\end{equation}
and
\begin{equation}
\label{eq:Whittle DOA deux sources avec et sans overlap r}
\lim_{n \rightarrow \infty}
n{\rm CRB}_r(\omega_i)
=\frac{1}{2\alpha_ir_i\int_{B_i}
\frac{mr_iR_{a_i}^2(f)}{1+2mr_iR_{a_i}(f)}df}.
\end{equation}
\end{result}

\begin{figure}[H]
\epsfxsize=0.7\hsize
\epsfysize=0.25\vsize
\centerline{\epsfbox{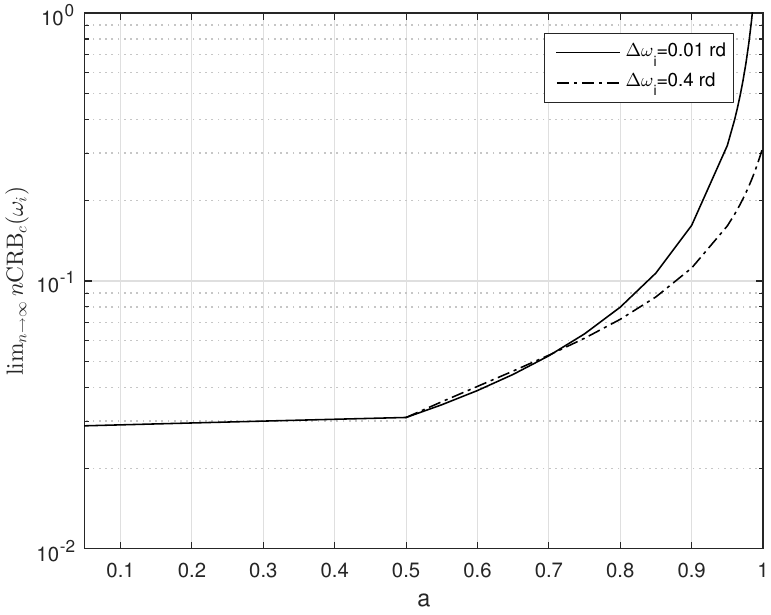}}
\vspace{-0.2cm}
\caption
{$\lim_{n \rightarrow \infty} {n{\rm CRB}_c(\omega_i)}$ as a function of $a$ for different values of $\Delta \omega$ for SNR = 0dB for two circular sources.}
\label{Fig 1}
\end{figure}

\begin{figure}[H]
\epsfxsize=0.7\hsize
\epsfysize=0.25\vsize
\centerline{\epsfbox{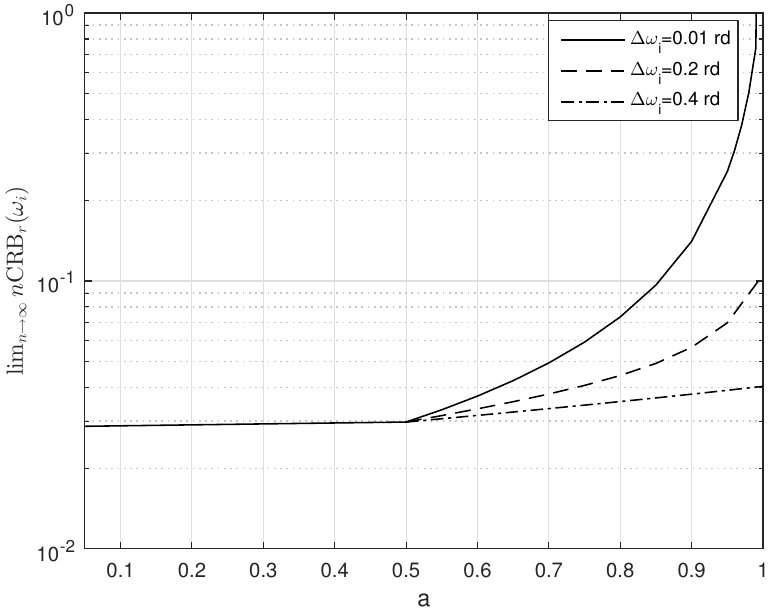}}
\vspace{-0.2cm}
\centerline{\small \footnotesize (a) $\Delta \phi =0.01$rd}

\vspace{0.2cm}

\epsfxsize=0.7\hsize
\epsfysize=0.25\vsize
\centerline{\epsfbox{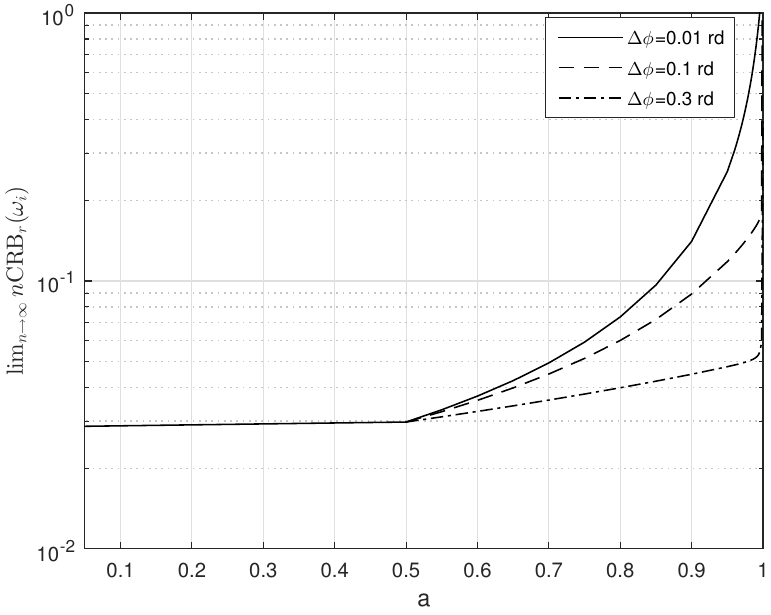}}
\vspace{-0.2cm}
\centerline{ \small \footnotesize (b) $\Delta \omega =0.01$rd}

\caption
{$\lim_{n \rightarrow \infty} {n{\rm CRB}_r(\omega_i)}$ as a function of $a$ for different values of $\Delta \omega$ and $\Delta \phi$ for SNR = 0dB for two rectilinear sources.}
\label{Fig 2}
\end{figure}

This result is illustrated in Fig. \ref{Fig 1} and Fig. \ref{Fig 2} for a circular and a rectilinear source, respectively. We
consider two equipowered sources with flat spectra $C_{s_1}(f)=R_{s_1}(f)=\frac{\sigma_{s}^2}{a}\mathds{1}_{[0,a]}(f)$ and $C_{s_2}(f)=R_{s_2}(f)=\frac{\sigma_{s}^2}{a}\mathds{1}_{[1-a,1]}(f)$  where $a \in (0,1]$ controls the degree of spectral overlap between the two sources. The sensor array is a uniform linear array (ULA) of $m=6$ elements with steering vectors ${\bf a}_i = (1, e^{i\omega_i}, \dots, e^{i(m-1)\omega_i})^T$, $i=1,2$. For this symmetric configuration, $\lim_{n \rightarrow \infty} n\, \mathrm{CRB}(\omega_1) = \lim_{n \rightarrow \infty} n\, \mathrm{CRB}(\omega_2)$, and we examine several values of the angular separation $\Delta\omega \pardef \omega_2 - \omega_1$ and phase difference $\Delta\phi \pardef \phi_2 - \phi_1$.

We verify that for non-overlapping spectra (i.e., $a \in (0, 1/2]$), the asymptotic CRB, $\lim_{n \rightarrow \infty} n\, \mathrm{CRB}(\omega_i)$, is exactly given by \eqref{eq:spectre plat c}-\eqref{eq:spectre plat r}, confirming the decoupling property established in Result~\ref{result:DOA 2 sources spectre sans overlap}.
Moreover, it can be observed that the CRB increases significantly as the spectral overlap increases (i.e., as $a$ increases beyond $1/2$) and as  $\Delta\omega$ decreases for circular sources [resp., as both $\Delta\omega$ and $\Delta\phi$ decrease for rectilinear sources]. This degradation in estimation performance is due to the increased difficulty in resolving closely spaced sources with overlapping spectral content.

Finally, for the case of complete spectral overlap ($a = 1$), we observe that the numerical value of $\lim_{n \rightarrow \infty} n \cdot \mathrm{CRB}(\omega_i)$ computed from \eqref{eq:Whittle's formula} coincides exactly with the closed-form expression $n\, \mathrm{CRB}_c(\omega_i)$ given in \cite[eq.~(25)]{Abeida2022}, which was derived for the classical case of i.i.d. complex circular Gaussian signals ${\bf x}_k$. 
%

\vspace{0.3cm}

%
\subsection{Cyclostationary CGP}
\label{sec:Cyclostationary CGP}
%
In this subsection, we consider a single source $s_{k,1}$ that is cyclostationary with cycle period $p$, and is either circular or noncircular to deal with the
linearly modulated digital communication signals presented in Subsection
\ref{sec:Application to linearly modulated digital communication signals}.
As established in Subsections~\ref{sec:Asymptotic FIM rate for cyclostationary CGP with cycle period p}
and \ref{sec:Asymptotic FIM rate for stationary and cyclostationary complex-valued CGP}, the extension of the asymptotic frequency-domain FIM rate \eqref{eq:Whittle's formula} to complex-valued cyclostationary signals applies when $n$ is a multiple of $p$, the number of samples of the resulting stationary blocked process is $n/p$.

In the circular case, the blocked observation vector ${\bf x}'_k$ defined in \eqref{eq:yn cyclostat} can be written as
${\bf x}'_k = {\bf A}' {\bf s}'_{k,1} + {\bf n}'_k \in \mathbb{C}^{mp}$,
where ${\bf A}' \pardef\text{Blkdiag}({\bf a}_1, \ldots, {\bf a}_1) \in \mathbb{C}^{mp \times p}$ ($p$ identical blocks), ${\bf s}'_{k,1} = (s_{(k-1)p+1,1}, \ldots, s_{kp,1})^T \in \mathbb{C}^p$, and ${\bf n}'_k = ({\bf n}_{(k-1)p+1}^T, \ldots, {\bf n}_{kp}^T)^T \in \mathbb{C}^{mp}$. 
Consequently, the covariance sequence and spectrum of the blocked process are respectively given by
${\bf C}_{x'}(\ell)\pardef \e({\bf x}'_k{\bf x}_{k+\ell}^{'H})
={\bf A}'{\bf C}_{s'}(\ell){\bf A}^{'H}+\sigma_n^2\delta_{\ell,0}{\bf I}_{mp}$ and
${\bf C}_{x'}(f)={\bf A}'{\bf C}_{s'}(f){\bf A}^{'H}+\sigma_n^2{\bf I}_{mp}$,
where ${\bf C}_{s'}(\ell)$ and  ${\bf C}_{s'}(f)$ are the $p \times p$ covariance sequence and spectrum  of the blocked source process $({\bf s}'_{k,1})_{k \in \mathbb{Z}}$.
The poly-spectral matrix of the source $s_{k,1}$ as defined in \eqref{eq:cyclic spectral matrix} is denoted by ${\bf C}_{s_1}^{\rm cyc}(f)$.

In the noncircular case, the signal $\widetilde{\bf x}'_k$ defined from $\widetilde{\bf x}_k$ similarly as ${\bf x}'_k$ from ${\bf x}_k$
in \eqref{eq:yn cyclostat},  is written in the form $\widetilde{\bf x}'_k=\widetilde{\bf A}'\widetilde{\bf s}'_{k,1}+\widetilde{\bf n}'_k \in \mathbb{C}^{2mp}$ where
$\widetilde{\bf A}' \pardef
{\rm Blkdiag}({\bf a}_1,{\bf a}_1^*...,{\bf a}_1,{\bf a}_1^*)\in \mathbb{C}^{2mp \times 2p}$,
and $\widetilde{\bf s}'_{k,1}$ and $\widetilde{\bf n}'_k$ defined from $\widetilde{\bf s}_{k,1}$ and $\widetilde{\bf n}_k$, respectively
 similarly as ${\bf x}'_k$ from ${\bf x}_k$.
This model also implies that ${\bf C}_{\tilde{x}'}(k) =\widetilde{\bf A}'{\bf C}_{\tilde{s}'}(k)\widetilde{\bf A}^{'H}+\sigma_n^2\delta_{k,0}{\bf I}$
and ${\bf C}_{\tilde{x}'}(f) =\widetilde{\bf A}'{\bf C}_{\tilde{s}'}(f)\widetilde{\bf A}^{'H}+\sigma_n^2{\bf I}$, where ${\bf C}_{\tilde{s}'}(k)$ and ${\bf C}_{\tilde{s}'}(f)$ are the $2p \times 2p$ augmented covariance  sequence and  spectrum of $(\widetilde{\bf s}'_{k,1})_{k \in \mathbb{Z}}$.
The poly-spectral matrix of the augmented source $\tilde{s}_{k,1}$, as  defined in \eqref{eq:cyclic spectral matrix},  is denoted by ${\bf C}_{\tilde{s}_1}^{\rm cyc}(f)$.

The source power is defined as the average power over one cycle period, $\sigma_{s_1}^2 \pardef\frac{1}{p}\sum_{k=1}^p \e(|s_{k,1}|^2)$, and the SNR as $r_1 \pardef \sigma_{s_1}^2/\sigma_n^2$. Using the Wiener-Khinchin theorem for the blocked process, the instantaneous power can be expressed as $\e(|s_{k,1}|^2) = \int_0^1 [{\bf C}_{s'}(f)]_{k,k} \, df$. We assume that the source spectral matrices ${\bf C}_{s'}(f)$ and ${\bf C}_{\tilde{s}'}(f)$ are parameterized by the source power $\sigma_{s_1}^2$ and a real-valued spectral shape parameter $a_1$. Specifically, we write ${\bf C}_{s'}(f) = \sigma_{s_1}^2 {\bf C}_{a_1'}(f)$ and ${\bf C}_{\tilde{s}'}(f) = \sigma_{s_1}^2 {\bf C}_{\tilde{a}_1'}(f)$, where the normalized spectral shape matrices satisfy the trace conditions $\int_0^1 \mathrm{tr}({\bf C}_{a_1'}(f)) \, df = p$ and $\int_0^1 \mathrm{tr}({\bf C}_{\tilde{a}_1'}(f)) \, df = 2p$, respectively. Consequently, the corresponding poly-spectral (cyclic) matrices ${\bf C}_{s_1}^{\rm cyc}(f)$ and ${\bf C}_{\tilde{s}_1}^{\rm cyc}(f)$ inherit this parameterization, taking the form ${\bf C}_{s_1}^{\rm cyc}(f) = \sigma_{s_1}^2 {\bf C}_{a_1}^{\rm cyc}(f)$ and ${\bf C}_{\tilde{s}_1}^{\rm cyc}(f) = \sigma_{s_1}^2 {\bf C}_{\tilde{a}_1}^{\rm cyc}(f)$. The unknown parameter vector for both the circular and noncircular models is therefore $\boldsymbol{\theta} = (\omega_1, a_1, \sigma_{s_1}^2, \sigma_n^2)^T$. Applying the matrix inversion lemma twice to the observation spectra ${\bf C}_{x'}(f)$ and ${\bf C}_{\tilde{x}'}(f)$ yields the following result, the main steps of which are detailed in the Appendix:
\begin{result}
\label{result:DOA une source cyclostationnaire}
For a single cyclostationary source of period $p$, either circular or noncircular, and for $n$ a multiple of $p$, the DOA parameter $\omega_1$ is FIM-decoupled from the nuisance parameters $(a_1, \sigma_{s_1}^2, \sigma_n^2)$ in the asymptotic frequency-domain FIM rate \eqref{eq:Whittle's formula}.
Moreover, the DOA-related FIM entry reduces to the second term of  \eqref{eq:Whittle's formula}. Consequently it does not depend on the CG distribution, and the asymptotic CRB on $\omega_1$ is given by
\begin{equation}
\label{eq:Whittle DOA une source circulaire cyclostationnaire}
\lim_{n \rightarrow \infty}
n\, {\rm CRB}_c(\omega_1)
=
\frac{1}{\alpha_1r_1p\int_0^1
mr_1\tra[{\bf C}_{a_1}^{\rm cyc}(f)({\bf I}_p+mr_1p\, {\bf C}_{a_1}^{\rm cyc}(f)]^{-1}{\bf C}_{a_1}^{\rm cyc}(f)]df},
\end{equation}
for the circular case, and
\begin{equation}
\label{eq:Whittle DOA une source noncirculaire cyclostationnaire}
\lim_{n \rightarrow \infty}
n\, {\rm CRB}_{nc}(\omega_1)
=
\frac{2}{\alpha_1r_1p\int_0^1
mr_1\tra[ {\bf C}_{\tilde{a}_1}^{\rm cyc}(f)({\bf I}_p+mr_1p\,{\bf C}_{\tilde{a}_1}^{\rm cyc}(f))^{-1}{\bf C}_{\tilde{a}_1}^{\rm cyc}(f)]df},
\end{equation}
for the noncircular case.
\end{result}

Finally, we verify that for the stationary case ($p=1$), we have ${\bf C}_{a_1}^{\mathrm{cyc}}(f) = C_{a_1}(f)$. Consequently, \eqref{eq:Whittle DOA une source circulaire cyclostationnaire} naturally reduces to \eqref{eq:Whittle DOA une source circulaire}, recovering the classical result for a circular stationary source. 
But since noncircular sources are not necessarily rectilinear, \eqref{eq:Whittle DOA une source noncirculaire cyclostationnaire} do not reduce to \eqref{eq:Whittle DOA une source rectilinear} for $p=1$.
%
\subsection{Stationary and cyclostationary CGP with frequency offset}
\label{sec:Stationary and cyclostationary CGP with frequency offset}
%
For a single source $s_{k,1}$ that is circular or noncircular, stationary or cyclostationary, and affected by a 
frequency offset $f_1 \in (0,1)$, model  \eqref{eq:DOA model} becomes:
\begin{equation}
\label{eq:modele offset DOA}
{\bf x}_k
=
e^{i2\pi k f_1}
s_{k,1}
{\bf a}_{1}+{\bf n}_k.
\end{equation}
This model is equivalent to  the second-order  model \eqref{eq:almost cyclostationary} with ${\bf u}_k
\pardef
s_{k,1}
{\bf a}_{1}+{\bf n}'_k$ where ${\bf n}'_k \pardef e^{-i2\pi k f_1}{\bf n}_k$. Since $({\bf n}_k)_{k \in \mathbb{Z}}$ is circular, the phase-rotated noise $({\bf n}'_k)_{k \in \mathbb{Z}}$ has identical second-order statistics, so $({\bf u}_k)_{k \in \mathbb{Z}}$ inherits the same covariance and pseudo-covariance structure as $s_{k,1}{\bf a}_1 + {\bf n}_k$.

Consequently, Result \ref{result:almost cyclo Whittle formula} applies, and therefore, Results \ref{result:DOA une source stat}, \ref{result:DOA 2 sources spectre sans overlap} and \ref{result:DOA une source cyclostationnaire} giving
an interpretable closed-form expression of $\lim_{n \rightarrow \infty}
n\, {\rm CRB}_{c}(\omega_1)$, $\lim_{n \rightarrow \infty}
n\, {\rm CRB}_{r}(\omega_1)$ and $\lim_{n \rightarrow \infty}
n\,{\rm CRB}_{nc}(\omega_1)$, respectively, also apply.
Furthermore, for the specific case of a circular stationary  source, the following result is proven in the Appendix.
\begin{result}
\label{result:DOA une source stat avec offset}
For a circular stationary source following model \eqref{eq:modele offset DOA} with  frequency offset $f_1$, the asymptotic CRB on $f_1$ is given by
\begin{equation}
\label{eq:Whittle offset une source circulaire s}
\lim_{n \rightarrow \infty}
n\, {\rm CRB}_c(f_1)
=
\frac{1}{m^2r_1^2\int_0^1\frac{(\frac{\partial C_{a_1}(f)}{\partial f})^2}{(1+mr_1C_{a_1}(f))^2}df}
\end{equation}
\end{result}
Note that the integral in \eqref{eq:Whittle offset une source circulaire s} is invariant to the frequency shift $f_1$ (since the integration is over a full period $[0,1]$), so the CRB on $f_1$ does not depend on the actual value of the offset.
Two important observations follow from \eqref{eq:Whittle offset une source circulaire s}:
\begin{itemize}
    \item The frequency offset $f_1$ is unidentifiable when the source spectrum is flat ($C_{a_1}(f)$ constant), since $\partial C_{a_1}(f)/\partial f = 0$ yields an infinite CRB.
    \item The asymptotic CRB on $f_1$ is independent of the sensor array geometry, as the expression does not involve the geometric sensitivity parameter $\alpha_1$ or the steering vector ${\bf a}_1$.
\end{itemize}

\begin{figure}[H]
\epsfxsize=0.7\hsize
\epsfysize=0.25\vsize
\centerline{\epsfbox{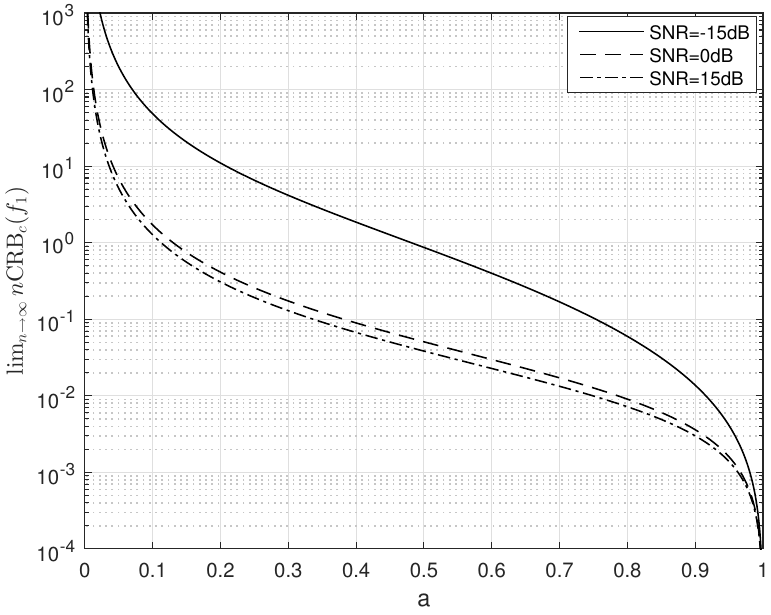}}
\vspace{-0.2cm}
\caption
{$\lim_{n \rightarrow \infty} {n{\rm CRB}_c(f_1)}$ as a function of $a$ for different values of SNR and $m=6$ for a circular AR(1) source.}
\label{Fig 3}
\end{figure}

This behavior is illustrated in Fig.~\ref{Fig 3}, which considers a circular stationary source with an AR(1) spectrum $C_{a_1}(f) = \frac{1-|a_1|^2}{|1-a_1 e^{-i2\pi f}|^2}$. As $|a_1| \to 0$, the spectrum becomes flat ($C_{a_1}(f) \to 1$), and the asymptotic CRB on $f_1$ tends to infinity, confirming that frequency offset estimation is impossible for spectrally white sources.
%
%
%
Note that in the case of two mutually uncorrelated circular stationary sources with spectra parameterized as $C_{s_1}(f) = \sigma_{s_1}^2 C_{a_1}(f)$ and $C_{s_2}(f) = \sigma_{s_2}^2 C_{a_2}(f)$, and each affected by its own frequency offset, model \eqref{eq:DOA model} becomes:
\begin{equation}
\label{eq:DOA deux sources avec offset}
{\bf x}_k
=
e^{i2\pi k f_1}
s_{k,1}
{\bf a}_{1}
+e^{i2\pi k f_2}
s_{k,2}
{\bf a}_{2}+{\bf n}_k.
\end{equation}
The process $({\bf x}_k)_{k \in \mathbb{Z}}$ remains circular and stationary, with spectral density matrix:
\begin{equation}
\label{eq:spectre DOA deux sources avec offset}
{\bf C}_x(f)
=
\sigma_{s_1}^2 C_{a_1}(f+f_1){\bf a}_1{\bf a}_1^H+\sigma_{s_2}^2 C_{a_2}(f+f_2){\bf a}_2{\bf a}_2^H
+\sigma_n^2{\bf I}_m.
\end{equation}
Thus, the complex circular extension of \eqref{eq:Whittle's formula} (as summarized in Table~1) applies with the full parameter vector $\boldsymbol{\theta} = (f_1, f_2, \omega_1, \omega_2, a_1, a_2, \sigma_{s_1}^2, \sigma_{s_2}^2, \sigma_n^2)^T$. Since these parameters are not decoupled in the FIM (both sources contribute to the same aggregate spectrum), it is not feasible to derive closed-form expressions for $\lim_{n \rightarrow \infty} n\, \mathrm{CRB}_c(\omega_i)$ from \eqref{eq:Whittle's formula} adapted to complex circular stationary processes.

\begin{figure}[H]
\epsfxsize=0.7\hsize
\epsfysize=0.25\vsize
\centerline{\epsfbox{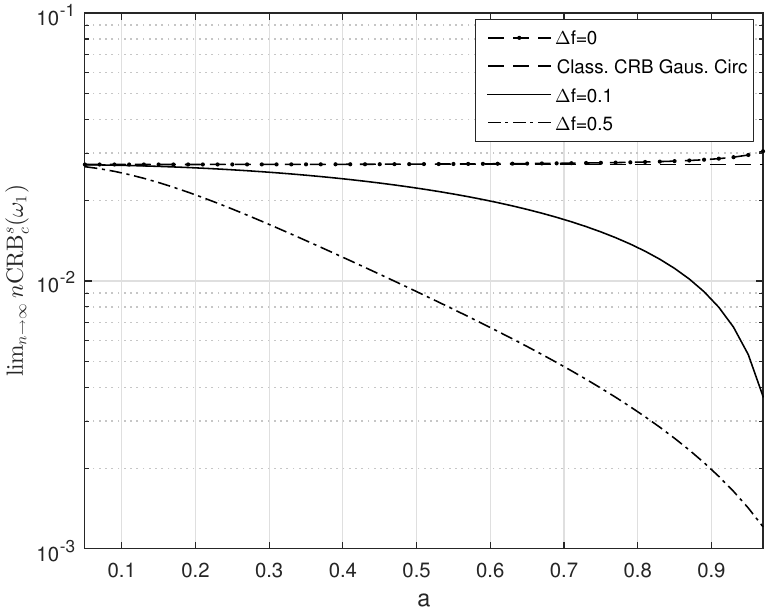}}
\vspace{-0.2cm}
\caption
{$\lim_{n \rightarrow \infty} {n{\rm CRB}_c(\omega_i)}$ as a function of $a$  for different values of $\Delta f$ for $\Delta \omega=0.2$rd and $r=15$dB.}
\label{Fig 4}
\end{figure}

\begin{figure}[H]
\epsfxsize=0.7\hsize
\epsfysize=0.25\vsize
\centerline{\epsfbox{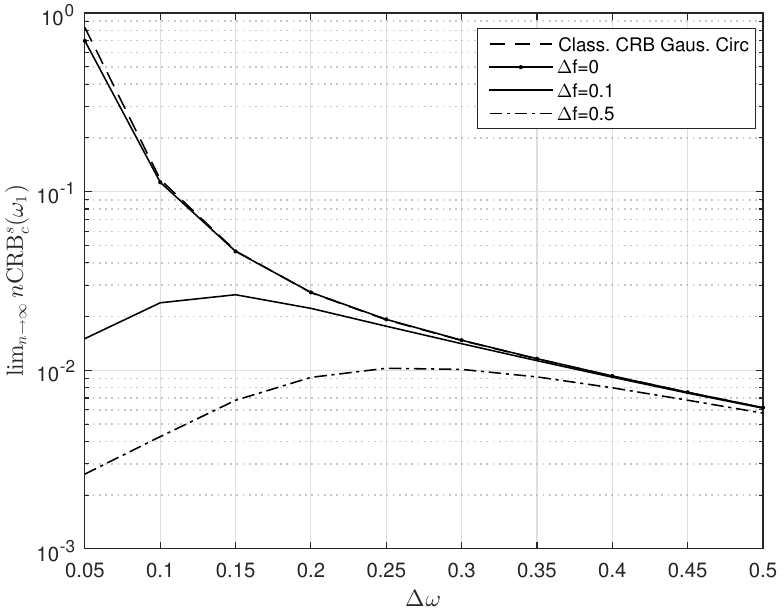}}
\vspace{-0.2cm}
\caption
{$\lim_{n \rightarrow \infty} {n{\rm CRB}_c(\omega_i)}$ as a function of $\Delta \omega$  for different values of $\Delta f$ for $a=0.5$ and $r=15$dB.}
\label{Fig 5}
\end{figure}

Nevertheless, numerical evaluation of \eqref{eq:Whittle's formula} permits investigation of the influence of the frequency offsets $f_1$ and $f_2$ on the DOA CRBs, which may be significantly affected by the source spectral characteristics. Consider two equipowered ($\sigma_s^2 \pardef \sigma_{s_1}^2 = \sigma_{s_2}^2$) uncorrelated sources with identical AR(1) spectra, where $a \pardef a_1 = a_2$, impinging on a ULA of $m=6$ sensors. We show numerically from \eqref{eq:Whittle's formula} that $\lim_{n \rightarrow \infty} n\, \mathrm{CRB}_c(\omega_1) = \lim_{n \rightarrow \infty} n\, \mathrm{CRB}_c(\omega_2)$, which depends only on the frequency offset difference $\Delta f \pardef f_1 - f_2$, the angular separation $\Delta\omega \pardef \omega_1 - \omega_2$, the AR(1) parameter $a$, and the SNR $r \pardef \sigma_s^2/\sigma_n^2$.
Figs.~\ref{Fig 4} and \ref{Fig 5} display $\lim_{n \rightarrow \infty} n\, \mathrm{CRB}_c(\omega_i)$ as a function of $a$ and $\Delta\omega$, respectively, for varying values of $\Delta f$. These figures also show the numerical values of $n\, \mathrm{CRB}_c(\omega_i)$ given by the closed-form expression \cite[eq.~(5)]{Abeida2022} for two independent circular Gaussian observations without frequency offset.
For $\Delta f = 0$ (i.e., identical frequency offsets), these figures clearly show that the CRB exhibits very low sensitivity to the spectral shape, reflecting the behavior observed in the single-source case. In contrast, for $\Delta f \neq 0$, the CRB is markedly sensitive to the spectral shape, exhibiting a significant reduction for strongly correlated samples (i.e., highly peaked spectra with $|a| \to 1$).

Furthermore, it is important to note that these asymptotic CRBs do not depend numerically on the CG distribution of the stationary observations. This is because the integral $\int_0^1 \mathrm{tr}\!\left({\bf C}_x^{-1}(f) \frac{\partial {\bf C}_x(f)}{\partial \omega_i}\right) df \approx 0$, so the CG-dependent correction term $c$ in \eqref{eq:Whittle's formula} has negligible effect on the asymptotic CRB. This property contrasts with the case of independent identically CG distributed observations, where $\mathrm{CRB}_c(\omega_i)$ is inversely proportional to $n\xi_{1,m}^c$ with $\xi_{1,m}^c \pardef \frac{\e[\mathcal{Q}_m^2 \phi_m^2(\mathcal{Q}_m)]}{m(m+1)} \leq 1$ \cite{Abeida2022}, and thus depends explicitly on the CG distribution even asymptotically.
%

%

%
%
Finally, it is worth noting that the numerical CRBs analyzed in \cite{Schell1992a} were derived under the assumption of stationary, rectilinear Gaussian sources with flat spectra. In that specific context, the effect of a carrier frequency offset manifests simply as a reduction in the effective source power across the full receiver bandwidth.
%
\section{Conclusion}
\label{sec:Conclusion}
%
This paper has presented comprehensive extensions of the asymptotic frequency-domain FIM rate, originally introduced in \cite{Delmas2026} for multidimensional stationary real-valued CGPs. First, we extended this result to purely cyclostationary real-valued CGPs. Subsequently, we generalized the formulation to encompass circular and noncircular complex-valued multidimensional stationary CGPs, as well as both purely and almost cyclostationary regimes. These theoretical advancements establish a unified framework for evaluating the asymptotic FIM rate across a broad and practically relevant class of stationary and cyclostationary CGPs. A primary advantage of these generalizations is the significant reduction in computational complexity when evaluating the CRB for parameters characterizing the mean and spectral structure of such processes. While the derived expressions are broadly applicable to diverse parameter-estimation problems in signal processing, we specifically applied them to investigate the influence of source spectral characteristics on the CRB for DOA estimation.  Our analysis reveals that while the temporal correlation (spectral shape) of the sources has a bounded impact on estimation accuracy, the presence of spectral offsets exerts a pronounced and systematic effect on the resulting DOA CRB. These findings underscore the critical importance of accurately modeling the spectral characteristics and potential frequency offsets of source signals when assessing the fundamental performance limits of DOA estimators. Future work will focus on extending this theoretical framework to multidimensional, spatiotemporally stationary spherically invariant random fields, thereby broadening its applicability to advanced signal and image processing applications.
%
\section*{Appendix}
\label{sec:Appendix}
\subsection*{Proof of Result \ref{result:almost cyclo Whittle formula}}
%
Let us firstly consider the case of a circularly complex-valued ${\bf u}_k$.
From \eqref{eq:almost cyclostationary}, we obtain:
${\bf y}_n={\bf \Delta}_n(\theta_0){\bf z}_n  \in \mathbb{C}^{mn}$
where ${\bf \Delta}_n(\theta_0)\pardef {\rm Blkdiag}(e^{i 2\pi \theta_0}{\bf I}_m,...,e^{i 2\pi n \theta_0}{\bf I}_m)$ is unitary.
It then follows that
 $\boldsymbol{\Sigma}_{{\bf y}_n}\pardef \e({\bf y}_n{\bf y}_n^H)
={\bf \Delta}_n(\theta_0)\boldsymbol{\Sigma}_{{\bf z}_n}{\bf \Delta}_n^H(\theta_0)$,
$\boldsymbol{\Sigma}_{{\bf y}_n}^{-1}
={\bf \Delta}_n(\theta_0)\boldsymbol{\Sigma}_{{\bf z}_n}^{-1}{\bf \Delta}_n^H(\theta_0)$,
 $\boldsymbol{\Sigma}'_{{\bf y}_n,0}
 ={\bf \Delta}'_n(\theta_0)\boldsymbol{\Sigma}_{{\bf z}_n}{\bf \Delta}_n^H(\theta_0)+{\bf \Delta}_n(\theta_0)\boldsymbol{\Sigma}_{{\bf z}_n}{\bf \Delta}_n^{'H}(\theta_0)$ and
  $\boldsymbol{\Sigma}'_{{\bf y}_n,\ell}
 ={\bf \Delta}_n(\theta_0)\boldsymbol{\Sigma}'_{{\bf z}_n,\ell}{\bf \Delta}_n^H(\theta_0)$, $\ell=1,..,q$ where
${\bf \Delta}'_n(\theta_0)\pardef \frac{d{\bf \Delta}_n(\theta_0)}{d \theta_0}
=i2\pi {\bf \Delta}_n{\bf \Delta}_n(\theta_0)$ with ${\bf \Delta}_n \pardef {\rm Blkdiag}({\bf I}_m,2{\bf I}_m...,n{\bf I}_m)$.
Plugging the expressions of $\boldsymbol{\Sigma}_{{\bf y}_n}^{-1}$,
 $\boldsymbol{\Sigma}'_{{\bf y}_n,k}$ into
$\tra(\boldsymbol{\Sigma}_{{\bf y}_n}^{-1}\boldsymbol{\Sigma}_{{\bf y}_n,k}^{'})$ and
$\tra(\boldsymbol{\Sigma}_{{\bf y}_n}^{-1}\boldsymbol{\Sigma}_{{\bf y}_n,k}^{'}\boldsymbol{\Sigma}_{{\bf y}_n}^{-1}\boldsymbol{\Sigma}_{{\bf y}_n,\ell}^{'})$, we get after simple algebraic manipulations, for $k,\ell=1,..,q$:
\begin{eqnarray}
\label{eq:trace 2}
\tra(\boldsymbol{\Sigma}_{{\bf y}_n}^{-1}\boldsymbol{\Sigma}_{{\bf y}_n,0}^{'})
&=&
i2\pi \tra(
{\bf \Delta}_n)
-i2\pi \tra(
{\bf \Delta}_n)={\bf 0},
\\
\label{eq:trace 4}
\tra(\boldsymbol{\Sigma}_{{\bf y}_n}^{-1}\boldsymbol{\Sigma}_{{\bf y}_n,0}^{'}\boldsymbol{\Sigma}_{{\bf y}_n}^{-1}\boldsymbol{\Sigma}_{{\bf y}_n,\ell}^{'})
&=&
i2\pi\tra(
 {\bf \Delta}_n
 \boldsymbol{\Sigma}_{{\bf z}_n}^{-1}
 \boldsymbol{\Sigma}_{{\bf z}_n,\ell}^{'})
 -i2\pi\tra(
 {\bf \Delta}_n
 \boldsymbol{\Sigma}_{{\bf z}_n}^{-1}
 \boldsymbol{\Sigma}_{{\bf z}_n,\ell}^{'})
= {\bf 0}.
\end{eqnarray}
This allows us to conclude from the expression of FIM that  $\theta_0$  is decoupled from $\{\theta_k\}_{k=1}^{q}$. Furthermore, the remaining terms utilized in the calculation of the FIM elements associated with $\{\theta_k\}_{k=1}^{q}$ can be expressed as follows:
\begin{eqnarray}
\label{eq:trace 2b}
\tra(\boldsymbol{\Sigma}_{{\bf y}_n}^{-1}\boldsymbol{\Sigma}_{{\bf y}_n,\ell}^{'})
&=&
\tra(\boldsymbol{\Sigma}_{{\bf z}_n}^{-1}\boldsymbol{\Sigma}_{{\bf z}_n,\ell}^{'})
\\
\label{eq:trace 4b}
\tra(\boldsymbol{\Sigma}_{{\bf y}_n}^{-1}\boldsymbol{\Sigma}_{{\bf y}_n,k}^{'}\boldsymbol{\Sigma}_{{\bf y}_n}^{-1}\boldsymbol{\Sigma}_{{\bf y}_n,\ell}^{'})
&=&
\tra(\boldsymbol{\Sigma}_{{\bf z}_n}^{-1}\boldsymbol{\Sigma}_{{\bf z}_n,k}^{'}\boldsymbol{\Sigma}_{{\bf z}_n}^{-1}\boldsymbol{\Sigma}_{{\bf z}_n,\ell}^{'}).
\end{eqnarray}

For noncircular complex-valued ${\bf u}_k$, the proof follows the same steps, using
$\widetilde{\bf y}_n=\widetilde{\bf \Delta}_n(\theta_0)\widetilde{\bf z}_n  \in \mathbb{C}^{2mn}$
where $\widetilde{\bf \Delta}_n(\theta_0)\pardef
{\rm Blkdiag}(e^{i 2\pi \theta_0}{\bf I}_m,e^{-i 2\pi \theta_0}{\bf I}_m...,e^{i 2\pi n \theta_0}{\bf I}_m,e^{-i 2\pi n \theta_0}{\bf I}_m)$ and
$\widetilde{\bf z}_n \pardef (\widetilde{\bf u}_{1}^T,...,\widetilde{\bf u}_{n}^T)^T$.
\hfill
\QED
\subsection*{Proof of Result \ref{result:DOA une source stat}}
In order to compute the extended version of  \eqref{eq:Whittle's formula} for complex circular-valued signals, it is necessary to evaluate the elements of 
${\rm FIM}_{{\bf y}_n}(\boldsymbol{\theta})$. This can be achieved by employing the following readily derived identities:

 ${\bf C}_{x}^{-1}(f)= \frac{1}{\sigma_n^2}{\bf I}- \frac{r_1 C_{a_1}(f)}{m\sigma_{s_1}^2 C_{a_1}(f)+\sigma_n^2}{\bf a}_1{\bf a}_1^H$,
$ \frac{\partial {\bf C}_{x}(f)}{\partial \omega_1}
=\sigma_{s_1}^2 C_{a_1}(f)[{\bf a}_1{\bf d}_1^H+{\bf d}_1{\bf a}_1^H]$,
$ \frac{\partial {\bf C}_{x}(f)}{\partial a_1}
= \sigma_{s_1}^2 \frac{\partial C_{a_1}(f)}{\partial a_1}{\bf a}_1{\bf a}_1^H$,
$ \frac{\partial {\bf C}_{x}(f)}{\partial \sigma_{s_1}^2}
=C_{a_1}(f){\bf a}_1{\bf a}_1^H$,
$ \frac{\partial {\bf C}_{x}(f)}{\partial \sigma_n^2}
={\bf I}$, $\|{\bf a}_1\|^2=m$ and
${\bf a}_1^H{\bf d}_1+{\bf d}_1^H{\bf a}_1=0$,
which, after simple algebraic manipulation, leads to
$\left[{\rm FIM}_{{\bf y}_n}(\boldsymbol{\theta})\right]_{\omega_1,a_1}
=\left[{\rm FIM}_{{\bf y}_n}(\boldsymbol{\theta})\right]_{\omega_1,\sigma_{s_1}^2}
=\left[{\rm FIM}_{{\bf y}_n}(\boldsymbol{\theta})\right]_{\omega_1,\sigma_n^2}
=0$ (the parameter $\omega_1$ and $(a_1,\sigma_{s_1}^2,\sigma_n^2)$ are decoupled in the ${\rm FIM}_{{\bf y}_n}(\boldsymbol{\theta})$) with $\tra({\bf C}_{x}^{-1}(f)  \frac{\partial {\bf C}_{x}(f)}{\partial \omega_1})=0$.
This implies:
\begin{equation}
\label{eq:Whittle's formula DOA c}
\lim_{n \rightarrow \infty}
[n{\rm CRB}(\omega_1)]^{-1}
=
\int_{0}^{1}\tra({\bf C}_{x}^{-1}(f)  \frac{\partial {\bf C}_{x}(f)}{\partial \omega_1} {\bf C}_{x}^{-1}(f) \frac{\partial {\bf C}_{x}(f)}{\partial \omega_1})df.
\end{equation}

Upon evaluating the trace of the product matrices in \eqref{eq:Whittle's formula DOA c} that depend on the aforementioned identities, we obtain \eqref{eq:Whittle DOA une source circulaire} through straightforward algebraic manipulation using the properties  $\tra([{\bf a}_1{\bf d}_1^H+{\bf d}_1{\bf a}_1^H][{\bf a}_1{\bf d}_1^H+{\bf d}_1{\bf a}_1^H])=m \alpha_1$ and
$\tra({\bf a}_1{\bf a}_1^H[{\bf a}_1{\bf d}_1^H+{\bf d}_1{\bf a}_1^H]{\bf a}_1{\bf a}_1^H[{\bf a}_1{\bf d}_1^H+{\bf d}_1{\bf a}_1^H])=0$.
\hfill
\QED

For the lower bound of ${\rm CRB}_c(\omega_1)$, \eqref{eq:min max c}
$\Leftrightarrow \int_0^1\frac{mr_1C^2_{a_1}(f)}{1+mr_1C_{a_1}(f)}df < 1
\Leftrightarrow
\int_0^1(C_{a_1}(f)-\frac{C_{a_1}(f)}{1+mr_1C_{a_1}(f)})df < 1$
which is true because  $\int_0^1C_{a_1}(f)df=1$ and $C_{a_1}(f)\geq0 \ \ \forall f$.

For the upper bound of ${\rm CRB}_c(\omega_1)$, \eqref{eq:min max c}
$\Leftrightarrow \int_0^1\frac{mr_1C^2_{a_1}(f)}{1+mr_1C_{a_1}(f)}df \ge \frac{mr_1}{1+mr_1}
\Leftrightarrow
1-\int_0^1\frac{C_{a_1}(f)}{1+mr_1C_{a_1}(f)})df\ge \frac{mr_1}{1+mr_1}
\Leftrightarrow
1-[\frac{1}{mr_1}-\frac{1}{mr_1}\int_0^1\frac{1}{1+mr_1C_{a_1}(f)}df]\ge \frac{mr_1}{1+mr_1}
\Leftrightarrow
\int_0^1\frac{1}{1+mr_1C_{a_1}(f)}df \ge \frac{1}{1+mr_1}$.
To prove this last inequality, we return to the minimization of an integral under constraint. This is achieved through the discretization of the Riemann integral with $c_k \pardef C_{a_1}(\frac{k}{n})$, $k=1,..,n$, after which we consider the following
optimization problem for arbitrary $n$:
\begin{equation}
\label{eq:optimization}
\min_{c_1,...c_n}\sum_{k=1}^n\frac{1}{1+mr_1c_k}
\ \  \mbox{subject to}\ \
\sum_{k=1}^n c_k=n,
\end{equation}
which is a standard convex optimization problem that can be solved by the Lagrange multiplier method \cite[chap. 5]{Boyd2009}.
Using the Lagrange function
$L(c_1,...,c_n,\lambda) = \sum_{k=1}^n\frac{1}{1+mr_1c_k}+\lambda(\sum_{k=1}^n c_k-n)$, it is straightforward to see that
$L(c_1,...,c_n,\lambda)$ is minimum for $c_k=1, i=1,...,n$. Consequently, $\int_0^1\frac{1}{1+mr_1C_{a_1}(f)}df$ under the constraint $\int_0^1C_{a_1}(f)df=1$  is minimum for $C_{a_1}(f)=1$.
The proofs for ${\rm CRB}_r(\omega_1)$ \eqref{eq:Whittle DOA une source rectilinear},\eqref{eq:min max r}
proceed in a similar manner.
\hfill
\QED
\subsection*{Proof of Result \ref{result:DOA 2 sources spectre sans overlap}}
%
For  non overlapping spectra $C_{s_1}(f)$ and $C_{s_2}(f)$, the inversion matrix lemma still applies and leads to
\begin{equation}
\label{eq:inversion lemma non overlap}
{\bf C}_{x}^{-1}(f)
=\left\{\begin{array}{ll}
 \frac{1}{\sigma_n^2}{\bf I}- \frac{r_i C_{a_i}(f)}{m\sigma_{s_i}^2 C_{a_i}(f)+\sigma_n^2}{\bf a}_i{\bf a}_i^H & \mbox{for}\ f \in B_i, i=1,2\\
 \frac{1}{\sigma_n^2}{\bf I} &  \mbox{for}\ f \in [0,1], f \notin B_1,  f \notin B_2.\\
\end{array}
\right.
\end{equation}
Using  $\frac{\partial {\bf C}_{x}(f)}{\partial \omega_i}
=\sigma_{s_i}^2 C_{a_i}(f)[{\bf a}_i{\bf d}_i^H+{\bf d}_i{\bf a}_i^H]$ for $f \in B_i$ and following the derivation steps in the proof of Result \ref{result:DOA une source stat}, we arrive at
 $\tra({\bf C}_{x}^{-1}(f)  \frac{\partial {\bf C}_{x}(f)}{\partial \omega_1})
\tra({\bf C}_{x}^{-1}(f)  \frac{\partial {\bf C}_{x}(f)}{\partial \omega_2})=0$ for $f\in [0,1]$
and
$\tra({\bf C}_{x}^{-1}(f)  \frac{\partial {\bf C}_{x}(f)}{\partial \omega_1}{\bf C}_{x}^{-1}(f)  \frac{\partial {\bf C}_{x}(f)}{\partial \omega_2})=0$ for $f\in [0,1]$.
Thus, $[{\rm FIM}_{{\bf y}_n}(\boldsymbol{\theta})]_{\omega_1,\omega_2}=0$.
Similarly to the single source case, the decoupling between $(\omega_1,\omega_2)$ and $(\sigma_{s_1}^2,\sigma_{s_2}^2,\sigma_{n}^2)$ is proven, and the application of \eqref{eq:Whittle's formula DOA c} to $\omega_1$ and $\omega_2$ applies.
This leads to the deduction of \eqref{eq:Whittle DOA deux sources avec et sans overlap c}.
The proofs for non-overlapping rectilinear sources
follow the same steps.
\hfill
\QED
\subsection*{Proof of Result \ref{result:DOA une source cyclostationnaire}}
%
The proof proceeds by expressing the frequency-domain FIM rate in terms of the spectrum of the stationary stacked source process ${\bf s}'_{k,1}$. 
Applying the matrix inversion lemma to ${\bf C}_{x'}(f) = {\bf A}'{\bf C}_{s'}(f){\bf A}^{'H}+\sigma_n^2{\bf I}_{mp}$ gives:
${\bf C}_{x'}^{-1}(f)
=\frac{1}{\sigma_n^2}({\bf I}_{mp}-{\bf A}'{\bf B}{\bf A}'^H)$ with ${\bf B} \pardef
\left[\sigma_n^2{\bf C}_{s'}^{-1}(f)+{\bf A}'^H{\bf A}' \right]^{-1}=r_1\left[{\bf I}_p+mr_1{\bf C}_{a'_1}(f) \right]^{-1}{\bf C}_{a'_1}(f)$.
Then using the derivatives 
$\frac{\partial {\bf C}_{x'}(f)}{\partial \omega_1} = {\bf D}'{\bf C}_{s'}(f){\bf A}^{'H} + {\bf A}'{\bf C}_{s'}(f){\bf D}^{'H}$, 
$\frac{\partial {\bf C}_{x'}(f)}{\partial a_1} = \sigma_{s_1}^2{\bf A}' \frac{\partial {\bf C}_{a_1'}(f)}{\partial a_1}{\bf A}^{'H}$,
$\frac{\partial {\bf C}_{x'}(f)}{\partial \sigma_{s_1}^2} = {\bf A}'{\bf C}_{a_1'}(f){\bf A}^{'H}$ and
$\frac{\partial {\bf C}_{x'}(f)}{\partial \sigma_n^2} = {\bf I}_{mp}$ with ${\bf D}' = \text{Blkdiag}({\bf d}_1, \ldots, {\bf d}_1)$,
 and the identities ${\bf A}'^H{\bf A}' =m {\bf I}_p$ and ${\bf A}'^H{\bf D}' +{\bf D}'^H{\bf A}' =({\bf a}_1^H{\bf d}_1 +{\bf d}_1^H{\bf a}_1){\bf I}_p={\bf 0}$
leads after simple algebraic manipulations to $\left[{\rm FIM}_{{\bf y}_n}(\boldsymbol{\theta})\right]_{\omega_1,a_1}
=\left[{\rm FIM}_{{\bf y}_n}(\boldsymbol{\theta})\right]_{\omega_1,\sigma_{s_1}^2}
=\left[{\rm FIM}_{{\bf y}_n}(\boldsymbol{\theta})\right]_{\omega_1,\sigma_n^2}
=0$ (the parameter $\omega_1$ and $(a_1,\sigma_{s_1}^2,\sigma_n^2)$ are decoupled in the ${\rm FIM}_{{\bf y}_n}(\boldsymbol{\theta})$) with $\tra({\bf C}_{x'}^{-1}(f)  \frac{\partial {\bf C}_{x'}(f)}{\partial \omega_1})=0$.
This implies from the extended version of  \eqref{eq:Whittle's formula} for complex circular-valued cyclostationary signals:
\begin{equation}
\label{eq:Whittle's formula DOA circular cyclo}
\lim_{n \rightarrow \infty}
[n{\rm CRB}(\omega_1)]^{-1}
=
\int_{0}^{1}\tra({\bf C}_{x'}^{-1}(f)  \frac{\partial {\bf C}_{x'}(f)}{\partial \omega_1} {\bf C}_{x'}^{-1}(f) \frac{\partial {\bf C}_{x'}(f)}{\partial \omega_1})df.
\end{equation}
Upon evaluating the trace of the product matrices in \eqref{eq:Whittle's formula DOA circular cyclo}, using the same calculations as for the stationary circular case in Result \ref{result:DOA une source stat}, we obtain 
\begin{equation}
\label{eq:Whittle DOA une source circulaire cyclostationnaire stacked}
\lim_{n \rightarrow \infty}
n{\rm CRB}_c(\omega_1)
=
\frac{p}{\alpha_1r_1\int_0^1
mr_1\tra[ {\bf C}_{a_1'}(f)({\bf I}_p\!+\!mr_1{\bf C}_{a_1'}(f))^{-1}{\bf C}_{a_1'}(f)]df}.
\end{equation}
Finally, substituting the unitary EVD ${\bf C}_{a_1'}(f) = p{\bf U}_p^H(f){\bf C}_{a_1}^{\rm cyc}(f){\bf U}_p(f)$ from \eqref{eq:unitarily similar} and exploiting the unitary invariance of the trace ($\tra[{\bf U}_p^H(f) {\bf X} {\bf U}_p(f)] = \tra({\bf X})$) directly yields \eqref{eq:Whittle DOA une source circulaire cyclostationnaire}.

The same methodology is employed to prove \eqref{eq:Whittle DOA une source noncirculaire cyclostationnaire} for cyclostationary noncircular sources, using the augmented matrices $\widetilde{\bf A}' \pardef
{\rm Blkdiag}({\bf a}_1,{\bf a}_1^*...,{\bf a}_1,{\bf a}_1^*)$ and $\widetilde{\bf D}' \pardef
{\rm Blkdiag}({\bf d}_1,{\bf d}_1^*...,{\bf d}_1,{\bf d}_1^*)$.
\hfill
\QED
\subsection*{Proof of Result  \ref{result:DOA une source stat avec offset}}
%
Using the decoupling between $f_1$ and $(\omega_1,a_1,\sigma_{s_1}^2,\sigma_n^2)$ in the FIM, as outlined in Result \ref{result:almost cyclo Whittle formula} and \eqref{eq:Whittle's formula} extended to circular complex-valued ${\bf x}_k$, we obtain:
\begin{equation}
\label{eq:Whittle's formula DOA}
\lim_{n \rightarrow \infty}
[n{\rm CRB}(f_1)]^{-1}
=
\int_{0}^{1}\tra({\bf C}_{x}^{-1}(f)\frac{\partial {\bf C}_{x}(f)}{\partial f_1} {\bf C}_{x}^{-1}(f)\frac{\partial {\bf C}_{x}(f)}{\partial f_1})df
+
2c
\left[\int_{0}^{1}\tra({\bf C}_{x}^{-1}(f) \frac{\partial {\bf C}_{x}(f)}{\partial f_1})df\right]^2,
\end{equation}
where
${\bf C}_x(f)
=
\sigma_{s_1}^2C_{a_1}(f+f_1){\bf a}_1{\bf a}_1^H
+\sigma_n^2{\bf I}$,
${\bf C}_{x}^{-1}(f)= \frac{1}{\sigma_n^2}{\bf I}- \frac{r_1 C_{a_1}(f+f_1)}{m\sigma_{s_1}^2 C_{a_1}(f+f_1)+\sigma_n^2}{\bf a}_1{\bf a}_1^H$
and
$\frac{\partial {\bf C}_{x}(f)}{\partial f_1}
=\sigma_{s_1}^2\frac{\partial C_{a_1}(f+f_1)}{\partial f}{\bf a}_1{\bf a}_1^H$ because
$\frac{\partial C_{a_1}(f+f_1)}{\partial f_1}=\frac{\partial C_{a_1}(f+f_1)}{\partial f}$.
Upon inserting the values of ${\bf C}_{x}^{-1}(f)$ and $\frac{\partial {\bf C}_{x}(f)}{\partial f_1}$ into \eqref{eq:Whittle's formula DOA} and performing the requisite algebraic manipulations, we obtain
\begin{eqnarray}
\nonumber
\lim_{n \rightarrow \infty}
[n{\rm CRB}_c^s(f_1)]^{-1}
&=&
m^2r_1^2\int_0^1\frac{(\frac{\partial C_{a_1}(f+f_1)}{\partial f})^2}{(1+mr_1C_{a_1}(f+f_1))^2}df
\\
\label{eq:Whittle's formula DOA bis}
&+&
2c
\left[\int_{0}^{1}
\frac{mr_1^2}{1+mr_1C_{a_1}(f+f_1)}
\frac{\partial C_{a_1}(f+f_1)}{\partial f}
df\right]^2.
\end{eqnarray}
Using the periodicity of $C_{a_1}(f+f_1)$ in $f$ of period unity, the first term of \eqref{eq:Whittle's formula DOA bis} reduces to
\eqref{eq:Whittle offset une source circulaire s} and the second integral to
$r_1
\left[\ln \left(1+mr_1C_{a_1}(f+f_1) \right)\right]_{0}^{1}=0$.
\hfill
\QED
\section*{Acknowledgement}
Habti Abeida would like to acknowledge the Deanship of Graduate Studies
and Scientific Research, Taif University, for funding this work.
%

\end{document}